\documentclass[aps, prd,nofootinbib,superscriptaddress,reprint,onecolumn,notitlepage,11pt,twoside]{revtex4-1}
\usepackage[includeheadfoot,paperheight = 269mm,paperwidth = 190mm, top=18mm, left = 28mm, right=18mm, bottom=22mm]{geometry} 
\usepackage{graphicx}
\usepackage{amssymb}
\usepackage{amsmath}
\usepackage{fancyhdr}
\usepackage[british]{babel}
\usepackage{natbib}
\usepackage{hyperref}
\newcounter{contribution}
\begin{document}

\title{Physics of the Cosmic Microwave Background Radiation}

\author{David Wands\footnote{E-mail: david.wands@port.ac.uk}}
\affiliation{Institute of Cosmology and Gravitation, University of Portsmouth, Dennis Sciama Building, Burnaby Road, Portsmouth PO1 3FX, United Kingdom}

\author{Oliver F. Piattella\footnote{E-mail: oliver.piattella@pq.cnpq.br}}
\affiliation{Universidade Federal do Esp\'{\i}rito Santo, Departamento de F\'{\i}sica, Av. Fernando Ferrari, 514, Campus de Goiabeiras, CEP 29075-910, Vit\'oria, Esp\'{\i}rito Santo, Brazil}

\author{Luciano Casarini\footnote{E-mail: casarini.astro@gmail.com}}
\affiliation{Universidade Federal do Esp\'{\i}rito Santo, Departamento de F\'{\i}sica, Av. Fernando Ferrari, 514, Campus de Goiabeiras, CEP 29075-910, Vit\'oria, Esp\'{\i}rito Santo, Brazil}

\begin{abstract}
The cosmic microwave background (CMB) radiation provides a remarkable window onto the early universe, revealing its composition and structure. In these lectures we review and discuss the physics underlying the main features of the CMB.
\end{abstract}

\maketitle

\section{Introduction}

The cosmic microwave background (CMB) radiation provides a remarkable window onto the early universe, revealing its composition and structure. It is a relic, thermal radiation from a hot dense phase in the early evolution of our Universe which has now been cooled by the cosmic expansion to just three degrees above absolute zero. Its existence had been predicted in the 1940s by Alpher and Gamow \cite{Alpher:1948ve, Alpher:2014hea} and its discovery by Arno Penzias and Robert Wilson at Bell Labs in New Jersey, announced in 1965 \cite{Penzias:1965wn} was convincing evidence for most astronomers that the cosmos we see today emerged from a Hot Big Bang more than 10 billion years ago.

Since its discovery, many experiments have been performed to observe the CMB radiation at different frequencies, directions and polarisations, mostly with ground- and balloon-based detectors. 
These have established the remarkable uniformity of the CMB radiation, at a temperature of 2.7 Kelvin in all directions, with a small $\pm 3.3$~mK dipole due to the Doppler shift from our local motion (at 1 million kilometres  per hour) with respect to this cosmic background. 

However, the study of the CMB has been transformed over the last twenty years by three pivotal satellite experiments.
The first of these was the Cosmic Background Explorer (CoBE), launched by NASA in 1990~\cite{Smoot:1992td, Mather:1993ij}. It confirmed the black body spectrum with an astonishing precision, with deviations less than 50 parts per million~\cite{Fixsen:1996nj}. 
And in 1992 CoBE reported the detection of statistically significant temperature anisotropies in the CMB, at the level of
$\pm 30$ $\mu$K on 10 degree scales~\cite{Smoot:1992td}. 
COBE was succeeded by the Wilkinson Microwave Anisotropy Probe (WMAP) satellite, launched by NASA in 2001, which produced full sky maps in five frequencies (from 23 to 94 GHz) mapping the temperature anisotropies to sub-degree scales and determining the CMB polarisation on large angular scales for the first time.
The Planck satellite, launched by ESA in 2009, sets the current state of the art with nine separate frequency channels, measuring temperature fluctuations  to a millionth of a degree at an angular resolution down to 5 arc-minutes.
Planck intermediate data was released in 2013~\cite{Ade:2013sjv}.\footnote{After these lectures were given, full-mission data was released in 2015~\cite{Adam:2015rua}, with final polarisation data still to come.}

These lectures draw upon the excellent reviews of CMB physics by Hu and Dodelson \cite{Hu:2001bc, Hu:2008hd, WayneHuwebpage}, Komatsu \cite{Komatsuwebpage} and Crittenden \cite{Crittendenwebpage}. 
We also refer the reader to comprehensive reviews on cosmological perturbations by Mukhanov {\em et al.} \cite{Mukhanov:1990me} and Malik and Wands \cite{Malik:2008im}. 
Useful textbooks are those of Peebles~\cite{Peebles:1994xt}, Dodelson~\cite{Dodelson:2003ft}, Mukhanov~\cite{Mukhanov:2005sc} and Weinberg~\cite{Weinberg:2008zzc}. 
Throughout this chapter we will use natural units such that $\hbar=k_B=c = 1$.


\section{Background cosmology and the Hot Big Bang model}

We start by recalling the mathematical framework describing the expansion of the universe and the Hot Big Bang. Much of modern cosmology is based on general relativity and the framework of Friedmann, Lema\^itre, Robertson and Walker in the 1920's and 30's \cite{Friedmann:1924bb, Lemaitre:1927zz, Robertson:1935zz}, and Hubble's discovery of the expansion of the universe \cite{Hubble:1929ig}.
We can ``slice'' four-dimensional spacetime into expanding three-dimensional space at each cosmic time, $t$, with a uniform matter density and spatial curvature. Requiring spatial homogeneity and isotropy at each cosmic time is known as the cosmological principle, which picks out the following space-time metric:
\begin{equation}\label{friedmmetr}
ds^2 = -dt^2 + a^2(t)\left[\frac{dr^2}{1 - \kappa r^2} + r^2d\Omega^2\right]
\;,
\end{equation}
where $a(t)$ is the scale factor and $\kappa$ is the curvature of the maximally symmetric spatial slices, and we chose spherical coordinates with infinitesimal solid angle $d\Omega^2$. 
We will sometimes find it convenient to use conformal time, $\eta$, where $dt=ad\eta$ and the line element takes the form
\begin{equation}\label{friedmmetr2}
ds^2  = a^2(\eta)\left[-d\eta^2 + \frac{dr^2}{1 - \kappa r^2} + r^2d\Omega^2\right]\;,
\end{equation}

The Hubble expansion rate is defined as $H \equiv \dot a/ a$, where a dot denotes a derivative with respect to cosmic time $t$. 
The present value of $H$ is called the Hubble constant and denoted as $H_0$.
The value of $H_0$ is often given in the form
\begin{equation}
\label{H0}
H_0 = 100\ h\ {\rm km\ s}^{-1}\ {\rm Mpc}^{-1} \,.
\end{equation}
Using the energy constraint, from Einstein's equations of general relativity, one gets the Friedmann equation for the Hubble expansion
\begin{equation}\label{Friedeq}
H^2 = \frac{8\pi G}{3}\rho + \frac{\Lambda}{3} - \frac{\kappa}{a^2}\;,
\end{equation}
where we introduce the cosmological constant, $\Lambda$, and $\rho$, the energy density. The latter includes electrons, baryons (protons, neutrons and atomic nuclei), radiation (photons and neutrinos) and dark matter (non-baryonic massive particles, non-relativistic by the present day).

Dividing through by $H^2$, eq.~\eqref{Friedeq} can be cast in the following dimensionless form:
\begin{equation}
 \label{Friedmann}
1 =  \Omega + \Omega_\Lambda + \Omega_\kappa \;.
\end{equation}
where we define the relative contributions to the Hubble expansion
\begin{equation}
\Omega \equiv \frac{8\pi G\rho}{3H^3} \,, \quad
\Omega_\Lambda \equiv \frac{\Lambda}{3H^2} \,,\quad
\Omega_\kappa \equiv \frac{-\kappa}{a^2H^2} \,.
\end{equation}

In order to get a closed system of equations we must determine the evolution of the density, $\rho$ in eq.~(\ref{Friedeq}), as a function of the scale factor. For this we can use the continuity (energy conservation) equation 
\begin{equation}
\dot\rho = -3H (\rho+P) \,,
\end{equation}
plus an equation of state for the pressure, $P(\rho)$. We will be interested in three important cases:
\begin{itemize}
\item
{$\Omega=\Omega_r=1$, radiation domination}:
\begin{equation}
\label{raddoma}
P_r = \frac13\rho_r \quad \Rightarrow\quad
\rho_r \propto a^{-4} \quad \Rightarrow\quad
a\propto t^{1/2} \propto \eta\,.
\end{equation}
\item
{$\Omega=\Omega_m=1$, matter domination (Einstein-de Sitter)}: 
\begin{equation}
P_m = 0 \quad \Rightarrow\quad
\rho_m \propto a^{-3} \quad \Rightarrow\quad
a\propto t^{2/3} \propto \eta^2\,.
\end{equation}
\item
{$\Omega_\Lambda=1$, $\Lambda$ domination (de Sitter)}: 
\begin{equation}
a\propto e^{Ht} \propto (\eta_\infty-\eta)^{-1}\,.
\end{equation}
\end{itemize}

The CMB consists of photons which survive from an early, radiation-dominated, Hot Big Bang and have a small density with respect to non-relativistic matter today. Nonetheless the CMB holds a rich store of information about the history of our Universe, as we shall see. 
For example, recent observations of the CMB by Planck \cite{Ade:2013zuv} can be used to infer values for the above cosmological parameters at the present-day:
\begin{equation}
 h = 0.674 \pm 0.014\;, \quad \Omega_{0} = 0.314 \pm 0.020\;, \quad \Omega_{\Lambda 0} = 0.686 \pm 0.020\;, \quad \Omega_{\kappa 0} = -0.04 \pm 0.05\;.
\end{equation}
The data are consistent with a flat universe, $\kappa = 0$, which will be our working hypothesis hereafter. We see that the expansion today is dominated by a cosmological constant (or some form of matter which acts very much like a cosmological constant) but in the recent past it was dominated by non-relativistic matter, and before that by radiation.

\subsection{Black-body spectrum}

The CMB is observed to have a black-body spectrum characteristic of a thermal equilibrium distribution, consistent with the hypothesis that our Universe emerged from a hot, dense Big Bang.

Photons follow a null trajectory in the FLRW metric \eqref{friedmmetr2} such that
\begin{equation}
\frac{dx^i}{d\eta} = \hat{n}^i \,,
\end{equation}
where $\hat{n}^i$ is a unit 3-vector, $g_{ij}\hat{n}^i\hat{n}^j=1$.
The 3-momentum of a photon is $p^i=p\hat{n}^i$, where $p$ is the wavenumber (remembering that we are using units such that $\hbar=1$ and $c=1$, so that $p$ also describes the energy of a massless photon).

CMB photons have an isotropic Bose-Einstein distribution function with temperature $T$
\begin{equation}
 \label{BEdist}
 f(p) = \frac{1}{\exp(p/T) - 1}\;.
\end{equation}
Given this isotropic distribution, we can compute the number density of CMB photons
\begin{equation}
 n_\gamma = 2\int \frac{4\pi p^2dp}{(2\pi)^3} f(p) \approx \frac{2.4}{\pi^2}T^3\;,
\end{equation}
where the photons have $2$ independent polarisation states and $4\pi p^2dp$ is the volume of an infinitesimal shell in three-dimensional momentum-space.
Their energy density is
\begin{equation}
 \label{rhogamma}
 \rho_\gamma  = 2\int \frac{4\pi p^2dp}{(2\pi)^3} p f(p) = \frac{\pi^2}{15}T^4\;.
\end{equation}

However the CMB photons are no longer in equilibrium with the matter we see in the universe today. The photons are free to propagate through the universe after electrons and baryons have recombined into neutral atoms, so the black-body spectrum must be propagated to the present day from the early universe.
Freely propagating photons follow the geodesic equation in curved space-time
\begin{equation}
\label{geodesic}
\frac{dP^\mu}{d\lambda} + \Gamma^\mu_{\nu\sigma} P^\nu P^\sigma = 0 \,,
\end{equation}
where $\Gamma^\mu_{\nu\sigma}$ is the Christoffel symbol. We define the photon 4-momentum as $P^\mu = dx^\mu/d\lambda$, where $\lambda$ is an affine parameter, and the modulus-squared of the 3-momentum is $p^2 = g_{ij}P^iP^j$ where $g_{ij}$ is the spatial part of FLRW metric \eqref{friedmmetr}. 
From the geodesic equation in the conformal FLRW metric \eqref{friedmmetr2} we obtain
\begin{equation}
\frac{1}{p}\frac{dp}{d\eta} = -\frac{1}{a}\frac{da}{d\eta}\;.
\end{equation}
Integrating this up to the present we obtain the cosmological redshift of the photon momentum, defined as
\begin{equation}
 1 + z \equiv \frac{p}{p_0} = \frac{a_0}{a}\;.
\end{equation}
We can interpret this simply as the expansion of the universe stretching the wavelength of a photon, reducing (redshifting) its energy and momentum. 

Note that the form of the Bose-Einstein distribution (\ref{BEdist}) is preserved
\begin{equation}
 f(p) = \frac{1}{\exp(p/T) - 1} = \frac{1}{\exp(p_0/T_0) - 1}\;,
\end{equation}
where the temperature is also redshifted with the expansion
\begin{equation}
 1 + z = \frac{T}{T_0} \;.
\end{equation}
Thus we see that the energy density (\ref{rhogamma}) of the photons decreases as the universe expands
\begin{equation}
\rho_\gamma \propto a^{-4} \,.
\end{equation}
Although photon density is small is in the universe today, it dominated the hot, dense, early universe.

\subsection{Hot Big Bang}

At sufficiently high temperatures we expect all particles to be relativistic. If these particles interact and efficiently redistribute energy they will share the same thermal equilibrium temperature. To be relativistic we require $T \gg m$, i.e., the thermal energy is much larger than the rest mass of a given particle species. At this stage of the primordial universe we can write the energy density using the same form given in eq.~\eqref{rhogamma} for all the relativistic species:
\begin{equation}
\label{rhoT}
 \rho = g_{\rm eff} \frac{\pi^2}{30}T^4\;,
\end{equation}
where $g_{\rm eff}$ is the sum of the effective number of degrees of freedom. Each bosonic species in thermal equilibrium contributes one per spin state (e.g., photons contribute $+2$, corresponding to two polarisations), whereas each fermion contributes $7/8$ per spin state, due to the different statistics.\footnote
{If a species decouples from this thermal bath, but remains relativistic, it can contribute with a different temperature in the above equation. This is what happens for neutrinos. They decouple relativistically from the primordial soup, at $T \approx 1$ MeV and their temperature today is expected to be $(4/11)^{1/3}$ times that of the photons because photons are heated by $e^-$-$e^+$ annihilation.}

In a radiation-dominated universe (\ref{raddoma}) the time dependence of the scale factor is given by $a\propto t^{1/2}$ and thus from eq.~(\ref{Friedeq}) we have
\begin{equation}
\rho = \frac{3H^2}{8\pi G} = \frac{3}{32\pi G t^2}\;,
\end{equation}
so that from (\ref{rhoT}) time and temperature are related by
\begin{equation}
 t = \sqrt{\frac{3}{32\pi G}\frac{30}{g_{\rm eff}\pi^2}}\frac{1}{T^2}\;.
\end{equation}
Thus we have the simple, approximate temperature-time relation
\begin{equation}
\frac{t}{1\mbox{ sec}} \approx \frac{1}{\sqrt{g_{\rm eff}}}\left(\frac{1\mbox{ MeV}}{T}\right)^2\;.
\end{equation}

\subsection{Spectral distortions}

The black-body shape of the CMB spectrum is maintained at early times because of the high interaction rate of photons with the other particles of the primordial plasma. We can identify two principal scattering processes which contribute to maintaining an isotropic, equilibrium distribution:
\begin{itemize}
%
\item{\em Compton scattering}: scattering of photons and relativistic electrons, redistributing energy and momentum, conserving photon number
\[
e^- + \gamma \ \leftrightarrow \ e^- + \gamma \,.
\]
At low energies this reduces to {\em Thomson scattering}, i.e., elastic scattering of photons off non-relativistic electrons, exchanging momentum, but conserving photon energy and number.
\item{\em Double (radiative) Compton scattering}: scattering of photons and relativistic electrons, redistributing energy and momentum, and changing photon number
\[
e^- + \gamma \ \leftrightarrow \ e^- + \gamma + \gamma \,.
\]
\end{itemize}

Many processes in the early universe before the time of recombination could potentially lead to measurable distortions in the CMB spectrum, which might be measured with future missions. Particle annihilation or decay would heat the primordial plasma, and hence the photons, or even the evaporation of primordial black holes in the relevant mass range. Even the damping of small scale density variations in the primordial plasma due to photon diffusion can lead to deviations from an exact black-body spectrum.
For more detail about CMB spectral distortions and what might cause them, see \cite{Chluba:2011hw}.

Efficient Compton and double Compton scattering maintains a full thermal equilibrium spectrum above a redshift~\cite{Hu:2008hd}
\begin{equation}
z_{th} = 2\times 10^6 \left( \frac{\Omega_bh^2}{0.02} \right)^{-2/5} \,,
\end{equation}
where $\Omega_b h^2$ determines the density of baryons and hence (in an electrically neutral universe) electrons.

Below this redshift Compton scattering can still redistribute energy and momentum between photons and electrons, but double Compton scattering becomes inefficient. In the absence of double Compton scattering, interactions cannot create or remove photons from the plasma. 
Compton scattering still maintains a statistical equilibrium above redshift~\cite{Hu:2008hd}
\begin{equation}
 z_\mu = 5\times 10^{4}\left(\frac{\Omega_{b0}}{0.02}\right)^{-1/2}\;.
\end{equation}
Thus if additional energy is dumped into the primordial plasma below redshift $z_{th}$ the CMB photons acquire a statistical equilibrium distribution
\begin{equation}
 f(p) = \frac{1}{\exp[(p - \mu)/T] - 1}\;.
\end{equation}
with non-zero chemical potential $\mu$. This is known as a $\mu$-distortion in the CMB spectrum.
Limits from the COBE satellite give an upper limit on the size of such a distortion~\cite{Fixsen:1996nj}:
\begin{equation}
\frac{|\mu|}{T} < 9\times 10^{-5}\ {\rm at}\ 95\%\ {\rm CL} \,. 
\end{equation}

Below the redshift $z_\mu$ Compton scattering off relativistic electrons becomes inefficient.
High-energy electrons along the line of sight can still transfer energy to low-frequency photons via inverse Compton scattering, without reaching statistical equilibrium. This leads to a characteristic ``y-distortion'' where low energy photons are boosted to higher frequencies, leading to a deficit in the CMB intensity at low frequencies in the Rayleigh-Jeans region, equivalent to a temperature deficit
\begin{equation}
\left.\frac{\Delta T}{T}\right|_{p \ll T} = -2y\;.
\end{equation}
and an enhancement at high frequencies.
The Compton $y$-parameter is defined as the line-of-sight integral of the electron pressure
\begin{equation}
 y = \int\frac{T_e}{m_e}n_e\sigma_Tdl\;.
\end{equation}
where $n_e$ is the density of free electrons and $\sigma_T$ is the Thomson scattering cross-section, see Eq.~(\ref{sigmaT}) below.
Constraints from COBE/FIRAS give the upper limit~\cite{Fixsen:1996nj}
\begin{equation}
|y| < 1.5 \times 10^{-5}\ {\rm at}\ 95\%\ {\rm CL} \,.
\end{equation}
These constraints still rely on COBE observations, more than twenty years ago. 

An important source of $y$-distortions seen in specific directions in the CMB is the Sunyaev-Zeldovich effect \cite{Sunyaev:1970eu}, from hot cluster gas along the line of sight after recombination. The Planck satellite has now compiled a catalogue of 439 clusters detected in the Planck data via their SZ signal \cite{Ade:2015fva} with many more being detected by ground-based experiments such as the Atacama Cosmology Telescope \cite{Hasselfield:2013wf} and the South Pole Telescope \cite{Bleem:2014iim}.

\subsection{Tight-coupling and sudden recombination}

At low energies (much smaller than the electron rest mass) electrons and photons interact via Thomson scattering, whose cross-section is
\footnote{The full cross-section describing the process $e^- + \gamma \rightarrow e^- + \gamma$ is given by the Klein-Nishina formula \cite{1929ZPhy...52..853K}, 
which displays not only the dependence on the photon energy but also on its polarization and the scattering angle. Since the energies involved in the recombination process are much smaller than the electron mass, we can safely use Thomson cross-section.}
\begin{equation}
\label{sigmaT}
 \sigma_{\rm T} = \frac{8\pi\alpha^2}{3m_{\rm e}^2} = 6.65\times 10^{-29}\mbox{ m}^2\;.
\end{equation}
The corresponding mean-free-path for photons associated with Thomson scattering is given by
\begin{equation}
\lambda_{\rm mfp} = \frac{1}{n_{\rm e}\sigma_{\rm T}}\;.
\end{equation}
Around $z\approx 1100$ the mean-free-path is approximately $2.5$~kpc, corresponding to a comoving scale of order $2.5$~Mpc at present~\cite{Hu:2008hd}.
On scales much larger than the mean-free-path, $\lambda\gg\lambda_{\rm mfp}$, the photons are tightly coupled to the electrons, while electrons are tightly coupled to protons through the Coulomb interaction.
In this regime, photons, electrons and protons can be treated as a single fluid with common 3-velocity, and isotropic pressure. 

The mean-free-path is time-dependent because the free-electron density, $n_{\rm e}$, is time-dependent. 
As the Universe cools down the capture of electrons by protons becomes efficient. 
As the wavelengths of photons are redshifted by the cosmic expansion, fewer photons have sufficient energy (the ionisation energy, $13.6$~eV) required to break the binding energy of an electron in a neutral hydrogen atom. Therefore, the density of free electrons, $n_e$, rapidly drops around $z\approx 1100$, leading to a rapid increase in the Thomson mean-free-path beyond the Hubble radius. 

This process is called decoupling, because photons no longer interact with electrons. It is also called recombination because this is the epoch when protons and electrons recombine to form hydrogen atoms. Recombination and decoupling are
practically simultaneous because the rapid drop in the density of free electrons due to recombination affects the Thomson scattering rate.
%
By solving the corresponding 
Boltzmann equation we see that recombination and decoupling occur 
at redshift~\cite{Hu:2008hd}
\begin{equation}
 1 + z_* = 1089\left(\frac{\Omega_mh^2}{0.14}\right)^{0.0105}\left(\frac{\Omega_bh^2}{0.024}\right)^{-0.028}\;.
\end{equation}
Note that this is some time after (but not long after) matter-radiation equality,
\begin{equation}
1+z_{\rm eq} = 3.4\times10^3\left(\frac{\Omega_m h^2}{0.14}\right) \,.
\end{equation}

Another way to define when recombination/decoupling takes place is via the Thomson optical depth
\begin{equation}
 \tau = \int_\eta^{\eta_0}n_e\sigma_Tdt\;,
\end{equation}
which represents the integrated scattering rate from a conformal time $\eta$ until today $\eta_0$, i.e., the average number of scattering events between these two times. 
The spatial hyper-surface of constant $\eta = \eta_*$, where $\eta_*$ is the conformal time corresponding to $\tau = 1$, is called the last-scattering surface.
Of course, recombination is not an instantaneous phenomenon, but it occurs sufficiently rapidly that a useful approximation on comoving scales greater than about $2.5$~Mpc is the so-called \textit{sudden recombination}, as if it really happened at a single instant, $\eta_*$. 


\section{CMB anisotropies}

Anisotropies observed in the CMB radiation are caused by inhomogeneities in the cosmological spacetime and matter distribution. Fortunately, these inhomogeneities are small (about $1$ part in $10^{4}$) with respect to the background homogenous energy density, thereby allowing us to use perturbation theory to model their behaviour. In the following we shall consider a linearly perturbed distribution.

We do not measure the plasma density directly, but rather anisotropies, in the CMB photon distribution function, $f \to \bar{f}+\delta f$.
At first order these can be described by a perturbation in the temperature of the Bose-Einstein distribution~(\ref{BEdist}), where
\begin{equation}\label{Thetafluct}
 T(\eta, \textbf{x},\hat{\textbf{p}}) =  \bar{T}(\eta)\left[1 + \Theta(\eta,\textbf{x},\hat{\textbf{p}})\right]\;,
\end{equation}
where $\hat{\textbf{p}}$ denotes the direction of the photon propagation.
The temperature fluctuation in the plasma is related to the photon density contrast via Eq.~(\ref{rhogamma}) as
\begin{equation}
 \Theta \equiv \frac{\delta T}{T} = \frac{1}{4}\frac{\delta\rho_\gamma}{\rho_\gamma} \equiv \frac14 \delta_\gamma\;.
\end{equation}

\subsection{Spherical harmonics}

Since we observe CMB on the celestial sphere, it is useful to expand $\Theta$ in spherical harmonics
\begin{equation}
\label{ThetaYlm}
 \Theta(\eta,\textbf{x},\hat{\textbf{p}}) = \sum_{\ell=0}^{\infty}\sum_{m=-\ell}^{\ell}a_{\ell m}(\eta,\textbf{x})Y_{\ell m}(\hat{\textbf{p}})\;.
\end{equation}
since the spherical harmonics form a complete orthonormal basis on the sphere
\begin{equation}\label{Ynorm}
\int d\Omega_n Y_{\ell m}(\hat{\textbf{n}}) Y^*_{\ell'm'}(\hat{\textbf{n}}) = \delta_{\ell\ell'}\delta_{mm'}\;.
\end{equation}
The coefficients $a_{lm}$ 
describe the 
temperature fluctuations at a given angular multipole $\ell$.
An isotropic distribution has an angular power spectrum $C_l$:
\begin{equation}
\label{Cldef}
 \langle a^*_{\ell m}a_{\ell'm'}\rangle = \delta_{\ell\ell'}\delta_{mm'}C_\ell\;.
\end{equation}
In this case the correlation between the temperatures in two directions on the CMB sky depends only on the angular distance between the two directions and not on the orientation of the arc which joins them.

For a fixed $\ell$, one has $2\ell+1$ different $a_{\ell m}$'s, i.e., $2\ell + 1$ independent estimates of the true $C_\ell$. The ``observed" $C_\ell^{\rm obs}$ corresponds to our best estimate of the true angular power spectrum:
\begin{equation}
 C_\ell^{\rm obs} \equiv \frac{1}{2\ell + 1}\sum_m (a^{\rm obs}_{\ell m})^* a^{\rm obs}_{\ell m}\;,
\end{equation}
i.e., it is an average over the observed multipole moments, $m$, at fixed $\ell$.
We define the cosmic variance as the expected error in our determination of the true power spectrum
\begin{equation}
 \left(\frac{\Delta C_\ell}{C_\ell}\right)^2_{\rm cosmic~variance} \equiv\left\langle\left(\frac{C_\ell - C_\ell^{\rm obs}}{C_\ell}\right)^2\right\rangle\;.
\end{equation}
Calculating the expectation in the above equation, with the help of eq.~\eqref{Cldef}, one obtains
\begin{equation}
 \left(\frac{\Delta C_\ell}{C_\ell}\right)_{\rm cosmic~variance} = \sqrt{\frac{2}{2\ell + 1}}\;.
\end{equation}
Thus at small multipoles, $\ell$, corresponding to very large angular scales, the cosmic variance is significant and represents the minimal uncertainty in estimating the true angular power spectrum given that we have only one realisation of the CMB sky.

\subsection{Last-scattering sphere}

Since most photons are last scattered at $\eta_*$, we will be mostly interested in their distribution, $\Theta(\eta,\textbf{x},\hat{\textbf{p}})$ in Eq.~(\ref{Thetafluct}),  at evaluated at recombination, i.e., at initial time $\eta=\eta_*$ and comoving displacement with respect to an observer at the origin, $\textbf{x}_*=-D_*\hat{\textbf{p}}$, where the comoving distance to last-scattering $D_*=\eta_0-\eta_*\simeq\eta_0$.
Then we propagate this photon distribution until today using the free-streaming equations, i.e., the collision-less Boltzmann equation for photons.

Adopting the sudden-recombination approximation, we assume that the photons are tightly coupled with an isotropic distribution up until last scattering.
\begin{equation}
\Theta_*(\hat{\textbf{p}}) = \Theta\left(\eta_*,\textbf{x}_* \right) \,.
\end{equation}
%
The CMB temperature varies across our sky due to the variation in the photon temperature across the last-scattering surface.

We can decompose this 3D CMB temperature field into Fourier modes 
\begin{equation}
\label{FourierTheta}
\Theta(\eta,\textbf{x}) = \frac{1}{(2\pi)^3}\int d^3\textbf{k} \, \Theta(\eta,\textbf{k}) \, e^{i\textbf{k}\cdot\textbf{x}}\;.
\end{equation}
Linear modes with different comoving wavevectors, $\textbf{k}$, then evolve independently at first order. 
%
We assume that these perturbations are stochastic quantities drawn from some distribution, 
which usually is assumed to be Gaussian. 

The expectation value of each mode is zero and its variance is the power spectrum
\begin{equation}\label{psthetadef}
\langle\Theta^*(\eta,\textbf{k}_1)\Theta(\eta,\textbf{k}_2)\rangle = (2\pi)^3\delta^3(\textbf{k}_1 + \textbf{k}_2)P_\Theta(k_1,\eta)\;.
\end{equation}
Note that $P_\Theta$ is function of the modulus of $\textbf{k}_1$ only, i.e., we assume statistical isotropy. 
%
The correlation function in real space is given by the Fourier transform of the power spectrum
\begin{equation}
\label{xirint}
\xi_\Theta(\textbf{r}) \equiv \langle\Theta(\eta,\textbf{x})\Theta(\eta,\textbf{x}+\textbf{r})\rangle = \frac{1}{(2\pi)^3}\int d^3\textbf{k}e^{i\textbf{k}\cdot\textbf{r}}P_\Theta(k)\;.
\end{equation}
%
Angle brackets denote the ensemble average. That is, one imagines different possible realizations of our universe. In theories such as inflation, where primordial fluctuations are quantum in their origin and then become effectively classical through an exponential phase of expansion, it is possible to predict the primordial form of the power spectrum. After that, it is evolved up until today using the classical equations of cosmological perturbation theory.
Thanks to the ergodic theorem, we can swap the ensemble average into a position 
average, see Appendix D of \cite{Weinberg:2008zzc}. 

Since $P_\Theta$ depends only on the modulus $k$, we can perform the angular integration in (\ref{xirint}) and find
\begin{equation}
 \xi_\Theta(r) = \frac{1}{2\pi^2}\int_0^\infty\frac{dk}{k}k^3P_\Theta(k)\frac{\sin kr}{kr}\;.
\end{equation}
From the above result, we can identify the dimensionless power spectrum
\begin{equation}
 \mathcal{P}_\Theta(k) \equiv \frac{k^3P_\Theta(k)}{2\pi^2}\;.
\end{equation}

We can decompose the temperature field on the last-scattering surface into spherical harmonics using the plane-wave expansion
\begin{equation}
e^{i\textbf{k}\cdot\textbf{r}} = 4\pi \sum_{l=0}^{\infty}\sum_{m=-l}^l i^\ell j_\ell(kr) Y^*_{\ell m}(\hat{\textbf{k}})Y_{\ell m}(\hat{\textbf{r}}) \,.
\end{equation}
where the spherical Bessel function $j_\ell(x)$ is defined in terms of the regular Bessel function $J_{\ell+1/2}(x)$ as $j_\ell(x)=(\pi/2x)^{1/2}J_{\ell+1/2}(x)$.
Substituting this expansion into (\ref{FourierTheta}) and comparing with (\ref{ThetaYlm}) evaluated at $\textbf{x}_*=-D_*\hat{\textbf{p}}$ we obtain the spherical harmonic coefficients
\begin{equation}
a_{\ell m} = \frac{i^\ell}{2\pi^2} \int d^3\textbf{k} \, \Theta(\eta_*,\textbf{k})\, j_l(kD_*)\, Y^*_{\ell m}(\hat{\textbf{k}})\;,
\end{equation}
and hence the angular power spectrum (\ref{Cldef}), by using eqs.~\eqref{Ynorm} and \eqref{psthetadef}, becomes:
\begin{equation}
 C_\ell = 4\pi\int_0^\infty\frac{dk}{k}\mathcal{P}_\Theta(k)j_\ell^2(kD_*)\;.
\end{equation}
The window function
\begin{equation}
 W_\ell(k) \equiv 4\pi j_\ell(kD_*)^2\;,
\end{equation}
peaks about $k = \ell/D_*$, so one obtains approximately that
\begin{equation}\label{largescalesCl}
 \frac{\ell(\ell + 1)}{2\pi}C_\ell \approx \mathcal{P}_\Theta(\ell/\eta_0)\;,
\end{equation}
by using $D_* \approx \eta_0$ and  the result
\begin{equation}
 \int_0^\infty\frac{dk}{k}j_\ell^2(k\eta_0) 
 = \frac{1}{2l(l + 1)}\;.
\end{equation}
This is the origin of the ubiquitous prefactor $l(l+1)$ in CMB spectrum plots. In order to obtain the full result one should include contributions from the metric perturbations and the dipole at recombination and the ISW effect, which we present in the following section. 


\section{Sachs-Wolfe formula}

In the previous section we discussed the basic quantities which describe the CMB temperature anisotropies at last-scattering, and in particular the angular power spectrum, $C_\ell$. In this section we link these to the observed temperature fluctuations including the effect of inhomogeneities in the metric and the density distribution of the matter content in the universe. We will derive the Sachs-Wolfe formula. In order to do this, we present the essential elements of relativistic cosmological perturbation theory, focusing on first-order fluctuations. The pioneering work in this field is due to Lifshitz \cite{Lifshitz:1945du} but we also refer the reader to more recent reviews, such as \cite{Malik:2008im}. 

\subsection{Metric perturbations}

The starting point for discussing cosmological perturbations is the perturbed FRLW metric \cite{Malik:2008im}
\begin{equation}
 ds^2 = a^2\left\{-(1 + 2A)d\eta^2 + 2\nabla_iBdx^id\eta + \left[(1+2C)\delta_{ij} + 2\nabla_i\nabla_jE\right]dx^idx^j\right\}\;,
\end{equation}
where $A$, $B$, $C$, and $E$ are scalar functions of the coordinates. In the above metric, we are considering only scalar perturbations, neglecting for now vector and tensor (gravitational wave) perturbations. Because of the tensorial nature of the metric, the above scalar functions change when changing the reference frame. It could happen that a reference frame exists in which $A = B = C = D = 0$. In this case then there are no metric perturbations, since we recover the original unperturbed FLRW metric. So, the fact of having four scalar functions of the coordinates in the above metric does not guarantee that we are actually dealing with cosmological perturbations, because the latter may be coordinate artifacts. This is the well-known \textit{gauge problem}. 

In order to know if we are really dealing with cosmological perturbations, a useful tool is to construct combinations of the above scalars which remain invariant under first order coordinate changes. There are three combinations independent of the spatial threading: $A$, $C$ and $\sigma \equiv E' - B$, where the prime denotes differentiation with respect to the conformal time $\eta$. There are then two combinations independent of time slicing, for example, the Bardeen potentials \cite{Bardeen:1980kt,Mukhanov:1990me,Malik:2008im}
\begin{equation}
\label{Bardeen}
\Psi \equiv A - \mathcal{H}\sigma - \sigma'\;, \qquad 
\Phi \equiv C - \mathcal{H}\sigma\;.
\end{equation}
In the above definition $\mathcal{H} \equiv a'/a$, is the conformal Hubble parameter, i.e. defined with respect to the conformal time. 

A particularly useful gauge is the conformal Newtonian gauge, where the metric becomes diagonal since the choice is $B = E = 0$. The Bardeen potentials (\ref{Bardeen}) can be identified with the metric perturbations $A$ and $C$ in this conformal Newtonian gauge (where $\sigma=0$).
The perturbed metric thus takes the form~\cite{Hu:2008hd}
\begin{equation}
\label{dscN}
 ds^2 = a^2\left\{-(1 + 2\Psi)d\eta^2 + (1+2\Phi)\delta_{ij}dx^idx^j\right\}\;.
\end{equation}
%
It can be shown by writing down explicitly the Einstein equations that their spatial traceless part depends on $\Phi + \Psi$. For example, the quadruple moment of the matter distribution acts as source of the spatial traceless part of the Einstein equation. In the tight coupling limit, there is no anisotropic stress because the high interaction rate of photons due to Thomson scattering establishes an isotropic distribution, which implies that $\Phi + \Psi = 0$.

One can construct other gauge-invariant variables, e.g., involving matter quantities, such as the density contrast and velocity potential in the conformal Newtonian gauge
\begin{equation}
 \delta \equiv \frac{\delta\rho - \rho'\sigma}{\rho}\;, \qquad V \equiv v + E'\;,
 \label{cNV}
\end{equation}
or the curvature perturbation 
\begin{equation}
\label{curvpert}
\zeta \equiv C - \frac{\mathcal{H}}{\rho'}\delta\rho 
\;,
\end{equation}
which can be identified with the metric perturbations $C$ in the uniform-density gauge.
This is a particularly useful variable on large scales since $\zeta$ is conserved for adiabatic perturbations on super-Hubble scales ($k\ll aH$) \cite{Wands:2000dp}. For example, simple slow-roll inflation models typically produce an approximately scale-invariant dimensionless power spectrum, $\mathcal{P}_\zeta(k)$, on large scales at the start of the radiation dominated era. Thus we will typically set initial conditions in terms of $\zeta$ and/or isocurvature perturbations.

Note that these different perturbation variables are not necessarily independent. For example we can express $\zeta$ in terms of the conformal Newtonian gauge quantities:
\begin{equation}
\zeta = \Phi - \frac{\mathcal{H}\rho}{\rho'}\delta \,.
\end{equation}

\subsection{Perturbed geodesics}

What is the form of a perturbed geodesics in the conformal Newtonian gauge (\ref{dscN})? By setting $ds^2 = 0$ for a null trajectory, we find the coordinate velocity of a photon
\begin{equation}
 \frac{dx^i}{d\eta} = (1 + \Psi - \Phi)\hat p^i\;,
\end{equation}
where $\hat p^i$ is a unit vector, $\delta_{ij}\hat p^i\hat p^j = 1$. Defining the 4-momentum as $P^\mu = dx^\mu/d\lambda$ and the modulus of the 3-momentum $p^2 = g_{ij}P^iP^j$, the perturbed geodesic equation (\ref{geodesic}) can be written as follows:
\begin{equation}\label{pertgeodeq}
 \frac{1}{p}\frac{dp}{d\eta} = -\left(\frac{1}{a}\frac{da}{d\eta} + \frac{\partial\Phi}{\partial\eta}\right) - \hat p^i\frac{\partial\Psi}{\partial x^i}\;.
\end{equation}
The term in parenthesis is the usual Hubble redshift corrected by the metric perturbation, which makes the expansion not homogeneous and isotropic, as it was in the background. The last term represents the gravitational blueshift or redshift experienced by a photon falling into or climbing out of a potential well. Introducing the total time derivative along the photon path, i.e.
\begin{equation}
\frac{d\Psi}{d\eta} = \frac{\partial\Psi}{\partial\eta} + \hat p^i\frac{\partial\Psi}{\partial x^i}\;,
\end{equation}
the geodesic equation \eqref{pertgeodeq} 
becomes
\begin{equation}
 \frac{1}{p}\frac{dp}{d\eta} = - \frac{1}{a}\frac{da}{d\eta} - \frac{d\Psi}{d\eta} 
 + \frac{\partial}{\partial\eta} \left( \Psi - \Phi \right)
 \;.
\end{equation}
This can be formally integrated along the photon trajectory from recombination, $\eta_*$, until today, $\eta_0$,
\begin{equation}
 \ln\left(\frac{p_0}{p_*} \right) = -\ln\left(\frac{a_0}{a_*} \right)  - \Psi_0 + \Psi_* + \int_{\eta_*}^{\eta_0}(\Psi' - \Phi')d\eta\;. 
\end{equation}
Splitting the momentum in a background part plus perturbation, i.e. $p \to p + \delta p$, one obtains
\begin{equation}
\label{perturbedmomentum}
 \left(\frac{\delta p}{p}\right)_0 = \left(\frac{\delta p}{p}\right)_* + \Psi_* - \Psi_0 + \int_{\eta_*}^{\eta_0}(\Psi' - \Phi')d\eta\;.
\end{equation}
This relative perturbation in the photon momentum causes a relative temperature fluctuation in the CMB, $\Theta = \delta p/p$. So, one sees that at recombination photons get a redshift escaping from over-densities on the last-scattering surface with a negative gravitational potential $\Psi_* $. This is part of the Sachs-Wolfe effect \cite{Sachs:1967er}. The integral term in (\ref{perturbedmomentum}), named the integrated Sachs-Wolfe effect, accounts for the time-dependence of the potentials along the line of sight from recombination until today. 

The observed temperature fluctuation includes a Doppler shift due to the relative peculiar velocity (in addition to the expansion) between the last-scattering surface and the observer
\begin{equation}
 \Theta = \frac{\delta p}{p} + \hat{\textbf{n}}\cdot \textbf{V}\;.
\end{equation}
The full Sachs-Wolfe formula is thus
\begin{equation}
\label{fullSW}
 \Theta_{\rm obs} = \frac{1}{4}\delta_{\gamma*} + \Psi_{*} - \hat{\textbf{n}}\cdot\textbf{V}_* + \int_{\eta_*}^{\eta_0}(\Psi' - \Phi')d\eta - \Psi_0 + \hat{\textbf{n}}\cdot\textbf{V}_{\rm obs}\;,
\end{equation}
where we have identified the relative momentum perturbation for photons on the last scattering surface with the radiation density contrast $\delta_\gamma=4\delta p/p$.
The first three terms represent the intrinsic Sachs-Wolfe effect (on the last-scattering surface) and the fourth the integrated one we already mentioned. $\Psi_0$ is the gravitational potential at the observer today and gives an undetectable correction to the monopole (that is, the solid-angle-averaged temperature). The last term is a dipole anisotropy induced by the observer's velocity.

\subsection{Adiabatic and isocurvature perturbations}

In order to evaluate the relative contribution of different terms in the Sachs-Wolfe formula (\ref{fullSW}) we need to determine the evolution of linear perturbations, in particular at the time of last scattering.

The behaviour  of the scalar perturbations previously introduced is given by the Einstein evolution equations (coming from the spatial part of the Einstein equations written in an arbitrary gauge):
\begin{eqnarray}
\label{Cevolution}
 C'' + 2\mathcal{H}C' - \mathcal{H}A' - (2\mathcal{H}' + \mathcal{H}^2)A &=& -4\pi G a^2\left(\delta P + \frac{2}{3}\nabla^2\Pi\right)\;,\\
 \sigma' + 2\mathcal{H}\sigma - C - A &=& 8\pi G a^2\Pi\;,
\end{eqnarray}
subject to the Einstein energy-momentum constraints (the time-time and time-space components of the Einstein equations):
\begin{eqnarray}
\label{energyconstraint}
 3\mathcal{H}(-C + \mathcal{H}A) + \nabla^2(C - \mathcal{H}\sigma) &=& -4\pi G a^2\delta\rho\;,\\
\label{mtmconstraint}
 -C' + \mathcal{H}A &=& -4\pi G a^2(\rho + P)(v + B)\;.
\end{eqnarray}
Finally, not independent from the above equations, we also have the energy and momentum conservation (continuity and Euler) equations:
\begin{eqnarray}
 \delta\rho' + 3\mathcal{H}(\delta\rho + \delta P) + 3(\rho + P)C' + (\rho + P)\nabla^2(v + E') &=& 0\;,\\
 (v + B)' + (1 - 3c_s^2)\mathcal{H}(v + B) + \phi + \frac{1}{\rho + P}\left(\delta P + \frac{2}{3}\nabla^2\Pi\right) &=& 0\;.
\end{eqnarray}
Note that the fluid quantities above introduced ($\delta\rho$, $\delta P$, etc.) refer to the total matter content, but if the components do not interact among themselves, these equations can also be considered individually for each one of the components that make up the balance of the cosmic energy budget. 

Exploiting the gauge freedom, we may consider the continuity equation written in the uniform density gauge, i.e. for $\delta\rho = 0$ where $\zeta=C$ in Eq.~(\ref{curvpert}):
\begin{equation}\label{unifgaugecons}
 3\mathcal{H}\delta P_{\rm nad} + 3(\rho + P)\zeta' + (\rho + P)\nabla^2 V = 0\;,
\end{equation}
where we identify the pressure perturbation with the non-adiabatic pressure in this uniform-density gauge, 
$\delta P_{\rm and}\equiv \delta P - (P'/\rho')\delta\rho$,
and the conformal Newtonian velocity $V$ was defined in Eq.~(\ref{cNV}).
Rearranging~\eqref{unifgaugecons} we have
\begin{equation}
 \zeta' = -\mathcal{H}\frac{\delta P_{\rm nad}}{\rho + P} - \frac{1}{3}\nabla^2V\;.
\end{equation}
For fluids with a barotropic equation of state, $P=P(\rho)$, we automatically have zero non-adiabatic pressure, $\delta P_{\rm nad} = 0$. Thus, on large scales, where the contribution from the divergence of the conformal Newtonian velocity, $\nabla^2V$, can be neglected, we have $\zeta$ being conserved. 

The same argument can be applied to any non-interacting barotropic fluids \cite{Lyth:2003im}. Thus, generalising the definition of the curvature perturbation~\eqref{curvpert}, we get conserved perturbations on large scales for radiation and matter, $\zeta_\gamma$ and $\zeta_m$. These can be written in terms of conformal Newtonian gauge quantities as
\begin{eqnarray}
\zeta_\gamma = \frac{\delta_\gamma}{4} + \Phi\;,\\
\zeta_m = \frac{\delta_m}{3} + \Phi\;.
\end{eqnarray}
Initial conditions are set up at sufficiently early times and on very large scales. Considering only radiation and matter, we have the total curvature and entropy perturbations
\begin{equation}
 \zeta = \frac{4\rho_\gamma\zeta_\gamma + 4\rho_m\zeta_m}{4\rho_\gamma + 3\rho_m}\;, \qquad S_m = 3(\zeta_m - \zeta_\gamma)\;.
\end{equation}
Adiabatic initial conditions are defined as:
\begin{equation}
 \zeta = \zeta_m = \zeta_\gamma = \mbox{constant}\;, \qquad S_m = 0\;.
\end{equation}
And the isocurvature initial conditions are defined as
\begin{equation}
 \zeta_\gamma = 0\;, \qquad S_m = 3\zeta_m = \mbox{constant}\;.
\end{equation}
A full treatment of the initial conditions requires neutrinos and baryons to be dealt with separately, giving rise to two more isocurvature density modes~\cite{Bucher:1999re}. 

The above constants are in general dependent on the scale. This dependence is set during an inflationary era and is thought to come from primordial quantum fluctuations. A slow time-dependence of the evolution during inflation leads to a weak scale-dependence of the dimensionless power spectrum. 
\begin{equation}
 \label{tilt}
n-1 \equiv \frac{d\ln {\mathcal{P}}_\zeta}{d\ln k} \approx 0 \,.
\end{equation}

The intrinsic Sachs-Wolfe effect (\ref{fullSW}) on large scales (neglecting the velocity $\textbf{V}_*$) can then be written as
\begin{equation}
 \frac{\delta T}{T} = \frac{1}{4}\delta_{\gamma*} + \Psi_* = \zeta_\gamma + 2\Psi_*\;.
\end{equation}
Let's now see how the evolution of the gravitational potential looks like, since we have seen its role in determining the Sachs-Wolfe effect. Consider a barotropic fluid with equation of state $p = w\rho$ and constant $w$. Neglecting anisotropic stresses, so that $\Phi = -\Psi$, the evolution equation is computed from the spatial trace of the Einstein equation \eqref{Cevolution} and reads:
\begin{equation}
 \Psi'' + 3(1 + w)\mathcal{H}\Psi' + w\nabla^2\Psi = 0\;.
\end{equation}
On large scales or during matter domination ($w=0$), where one can neglect the spatial gradient, we obtain a constant solution
\begin{equation}
 \Psi_0 = -\frac{3(1 + w)}{5 + 3w}\zeta\;,
\end{equation}
where we related this constant to $\zeta$ which we have already shown to be conserved on large scales and for adiabatic perturbations. 
This sets the initial conditions for the scales which subsequently enter the horizon. 

The evolution of the gravitational potential behaves in a two very different ways from the radiation dominated phase compared to the matter dominated era. In radiation, $w = 1/3$, for a comoving wavenumber $k$, we find 
\begin{equation}
 \Psi_k(\eta) = \zeta_\gamma \left[\frac{6}{(k\eta)^2}\cos\left(\frac{k\eta}{\sqrt{3}}\right) - \frac{6\sqrt{3}}{(k\eta)^3}\sin\left(\frac{k\eta}{\sqrt{3}}\right)\right]\;.
\end{equation}
For super horizon scales ($k\eta \ll 1$) the above solution tends to a constant: $\Psi_k \to -2\zeta_\gamma/3$.
But for sub horizon scales ($k\eta \to \infty$) the potential oscillates and decays, $\Psi_k \to 0$. Thus the growth of matter inhomogeneities will also be suppressed in this regime. 

In the matter dominated era ($w = 0$) the gravitational potential is constant at all scales
\begin{equation}
\Psi_k(\eta) = -3\zeta_m/5\,. 
\end{equation}
With this result, the intrinsic Sachs-Wolfe effect on large scales can be written as
\begin{equation}
 \frac{\delta T}{T} = \zeta_\gamma + 2\Psi_* = \zeta_\gamma - \frac{6}{5}\zeta_m\;,
\end{equation}
where we used the matter-dominated solution for the gravitational potential because, as previously noted, recombination takes place in this regime, $z_*<z_{\rm eq}$. For adiabatic perturbations ($S_m = 0$ and $\zeta_m = \zeta_\gamma$) the contribution is:
\begin{equation}
 \left.\frac{\delta T}{T}\right|_{\rm ad} = - \frac{1}{5}\zeta_\gamma = \frac{1}{3}\Psi_*\;,
\end{equation}
whereas for isocurvature perturbations ($\zeta_\gamma = 0$, $S_m = 3\zeta_m$):
\begin{equation}
 \left.\frac{\delta T}{T}\right|_{\rm iso} = - \frac{2}{5}S_m = 2\Psi_*\;.
\end{equation}
With these two formulas, the large-scale approximation \eqref{largescalesCl} can be written as follows:
\begin{equation}
 \frac{l(l + 1)}{2\pi}C_l \approx \frac{1}{25}\mathcal{P}_\zeta(l/\eta_0) + \frac{4}{25}\mathcal{P}_S(l/\eta_0) + \frac{4}{25}\mathcal{C}_{\zeta S}(l/\eta_0) \;,
\end{equation}
i.e. a contribution coming from the adiabatic perturbations, $\mathcal{P}_\zeta$, another from the isocurvature perturbations, $\mathcal{P}_S$, and their cross-correlation, $\mathcal{C}_{\zeta S}$ (recall that the power spectrum is a quadratic function of the perturbation variables).


\section{Acoustic oscillations}

One of the striking features of CMB power spectrum is the presence of acoustic oscillations, which originate from sound waves in the baryon-photon fluid at the time of last scattering. We present here the physics which lies behind this phenomenon, following the approach of Hu~\cite{Hu:2008hd}.

The coupled first-order energy and momentum conservation equations for radiation perturbations in conformal Newtonian gauge (neglecting for the moment the effect of baryons) are
\begin{eqnarray}
 \frac{1}{4}\delta_\gamma' &=& -\frac{1}{3}\nabla\cdot\textbf{V}_\gamma - \Phi'\;,\\
 \textbf{V}_\gamma' &=& -\frac{1}{4}\nabla\delta_\gamma - \nabla\Psi\;.
\end{eqnarray}
Eliminating $\textbf{V}_\gamma$ yields an oscillator equation:
\begin{equation}\label{osceq}
 \left(\frac{1}{4}\delta_\gamma + \Psi\right)'' - \frac{1}{3}\nabla^2\left(\frac{1}{4}\delta_\gamma + \Psi\right) = (\Psi - \Phi)''\;.
\end{equation}

\subsection{Matter era}

It is easy to solve this equation in the matter era, since $\Psi = -\Phi =
$ constant, therefore the right-hand-side vanishes and we get
\begin{equation}\label{acosc}
 \left(\frac{1}{4}\delta_\gamma + \Psi\right) = \left(\frac{1}{4}\delta_\gamma + \Psi\right)_0\cos(ks)\;,
\end{equation}
for $k \gg k_{\rm eq}$, where $k_{\rm eq}$ is the comoving wave-number of a scale which crosses the Hubble horizon at matter-radiation equivalence, i.e. $k_{\rm eq} = H_{\rm eq}a_{\rm eq}$.
The sound horizon for the relativistic fluid is given by
\begin{equation}
 s = \int_0^\eta c_sd\eta' = \frac{1}{\sqrt{3}}\eta\;.
\end{equation}
We shall see later that the presence of baryons affects the sound horizon since it modifies the speed of sound. 

\begin{figure}[ht]
\includegraphics[width=\columnwidth]{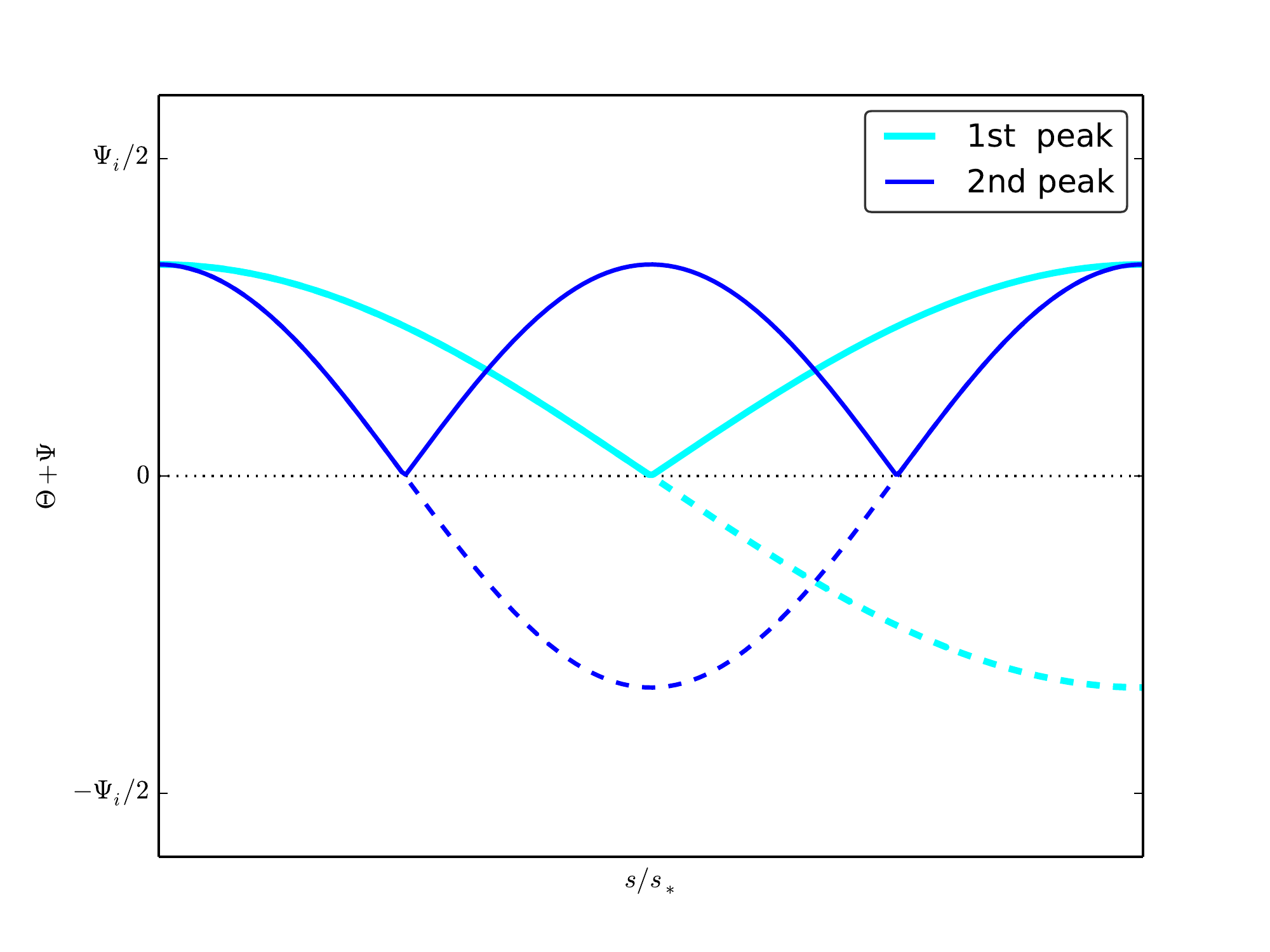}\\
\caption{Evolution of the effective temperature, eq.~\eqref{acosc}, and its absolute value (corresponding to the power peaks, solid lines) \cite{Hu:2008hd,Hu:2001bc}.}
\label{Fig1}
\end{figure}

At recombination, for adiabatic perturbations, the above solution can be written as
\begin{equation}\label{osceqsolmmater}
  \left(\frac{1}{4}\delta_\gamma + \Psi\right)_* = -\frac{1}{5}\zeta_\gamma\cos(ks_*)\;.
\end{equation}
We see in Fig.~\ref{Fig1} the characteristic oscillating behaviour of the CMB temperature fluctuations, known as acoustic oscillations. 
Note that the power spectrum corresponds to the square of the amplitude so that peaks in the power spectrum occur at maxima or minima of the amplitude.

The first peak takes place when the argument of the cosine is equal to $\pi$, i.e. the comoving acoustic scale, $\lambda_A$, is defined as
\begin{equation}
 k_As_* 
 = \pi \quad \Rightarrow \quad \lambda_A = \frac{2\eta_*}{\sqrt{3}}\;.
\end{equation}
Dividing by the comoving distance to recombination, we find the angular scale
\begin{equation}
 \theta_A = \frac{\lambda_A}{\eta_0 - \eta_*}\;,
 \end{equation}
which in the matter-dominated universe can be approximated as
 \begin{equation}
  \theta_A \approx \frac{\eta_*}{\eta_0} \approx z_*^{-1/2} \approx 2^{\circ}\;.
 \end{equation}
This is the angular scale of the particle horizon at recombination. It spans only 2 degrees in the sky, and yet we see a high degree of isotropy in the CMB sky on all angular scales. No causal process could lead to isotropy on scales separated by more than $2^\circ$ in the classical hot big bang because of the lack of casual connection on any large scales. This is the well-known horizon problem in the big bang model which is most clearly seen in the CMB, and to which inflation offers a solution.

\subsection{Radiation driving}


In the radiation dominated era, the gravitational potentials cannot be assumed to be constant and they rapidly oscillate and decay as each scale enters the horizon, i.e. $\Phi = -\Psi \to 0$ as $k\eta \to \infty$. Thus we can neglect the source term in the oscillator equation~\eqref{osceq} for $k\eta\gg1$
to obtain
\begin{equation}
 \left(\frac{1}{4}\delta_\gamma + \Psi\right) \approx \frac{1}{4}\delta_\gamma\propto\cos(ks)\;.
 \end{equation}

\begin{figure}[ht]
\includegraphics[width=\columnwidth]{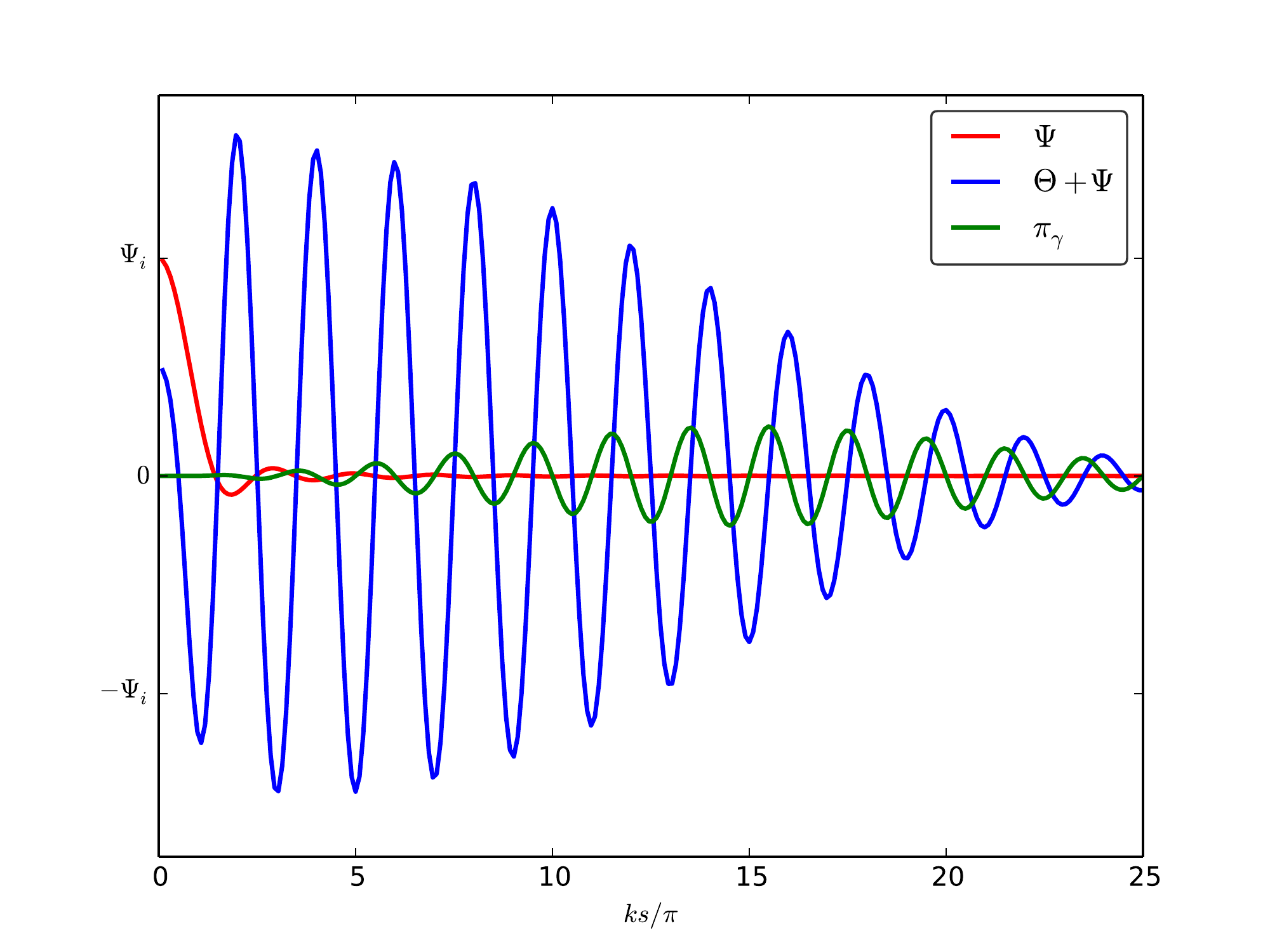}\\
\caption{Evolution, starting in the radiation era, of the gravitational potential, $\Psi$ (red), the effective temperature, $(1/4)\delta_\gamma+\Psi$ (blue), and the anisotropic pressure due to photon diffusion, $\pi_\gamma$ (green) \cite{Hu:2008hd}.}
\label{Fig3}
\end{figure}

Extrapolating forward to the matter era, we obtain the same oscillations for the radiation density as before, Eq.~(\ref{osceqsolmmater}), but with larger amplitude the potentials are negligible. For scales that enter the horizon well before matter-radiation equality, $k\gg k_{\rm eq}$, we find
\begin{equation}
 \left(\frac{1}{4}\delta_\gamma + \Psi\right)_* \approx -\zeta_\gamma\cos(ks_*)\;.
\end{equation}
Comparing with Eq.~(\ref{osceqsolmmater}), we find that the amplitude of oscillations for $k> k_{\rm eq}$ can be five times larger for low matter density, as this delays matter-domination, which causes the gravitational potential to decay on sub-horizon scales during the radiation era, see Figure~\ref{Fig3}.

\subsection{Baryon loading}

What happens if we include the effect of baryons which are strongly coupled to the electrons, and hence photons, before recombination? Intuitively, since baryons are massive particles, they would slow down the oscillations. 

We define the coupled baryon-photon velocity, $V_{b\gamma}$, using
\begin{equation}
 (\rho_\gamma + P_\gamma)\textbf{V}_\gamma + (\rho_b + P_b)\textbf{V}_b = (1 + R)(\rho_\gamma + P_\gamma)\textbf{V}_{b\gamma}\;,
\end{equation}
where the baryon-to-photon ratio is given by
\begin{equation}
R \equiv \frac{\rho_b + P_b}{\rho_\gamma + P_\gamma} = \frac{3}{4}\frac{\rho_b}{\rho_\gamma}\;.
\end{equation}
In the tight-coupling limit, $\textbf{V}_{b\gamma}= \textbf{V}_\gamma = \textbf{V}_b$. 

Using the coupled energy and momentum conservation equations we can write
\begin{eqnarray}
 \frac{1}{4}\delta_\gamma' &=& -\frac{1}{3}\nabla\cdot\textbf{V}_{b\gamma} - \Phi'\;,\\
 \left[(1 + R)\textbf{V}_{b\gamma}\right]' &=& -\frac{1}{4}\nabla\delta_\gamma - (1 + R)\nabla\Psi\;.
 \end{eqnarray}
Neglecting the time-variation of $\Psi$ and $R$ (i.e., $\Psi' \approx R' \approx 0$, which is a reasonable approximation) one gets the following oscillator equation:
 \begin{equation}
  \left[\frac{1}{4}\delta_\gamma + (1 + R)\Psi\right]'' - \frac{1}{3(1 + R)}\nabla^2\left[\frac{1}{4}\delta_\gamma + (1 + R)\Psi\right] \approx 0\;.
 \end{equation}
 Comparing with the earlier oscillator equation neglecting baryons, (\ref{osceq}), we see that the adiabatic speed of sound (the term multiplying the Laplacian is the square of the sound speed), and now is reduced by the presence of baryons by a factor $(1 + R)$. 
 The solution for the above equation is:
 \begin{equation}
  \left[\frac{1}{4}\delta_\gamma + (1 + R)\Psi\right]  = \left[\frac{1}{4}\delta_\gamma + (1 + R)\Psi\right] _0\cos(ks)\;,
 \end{equation}
 where now the sound horizon is 
 \begin{equation}
  s \equiv \int c_sd\eta = \int \frac{d\eta}{3(1+R)}\;.
 \end{equation}
 So the matter-era solution for $k \gg k_{\rm eq}$, \eqref{osceqsolmmater}, corrected by the presence of baryons is now:
 \begin{equation}
  \left(\frac{1}{4}\delta_\gamma + \Psi\right)_* \approx \left[3R - (1 + 3R)\cos(ks_*)\right]\frac{1}{5}\zeta_\gamma\;.
 \end{equation}
The oscillations are now asymmetric: compression and rarefaction are no longer symmetric because baryons tend to collapse under their own gravity, more than they become more rarefied.

 \begin{figure}[ht]
\includegraphics[width=\columnwidth]{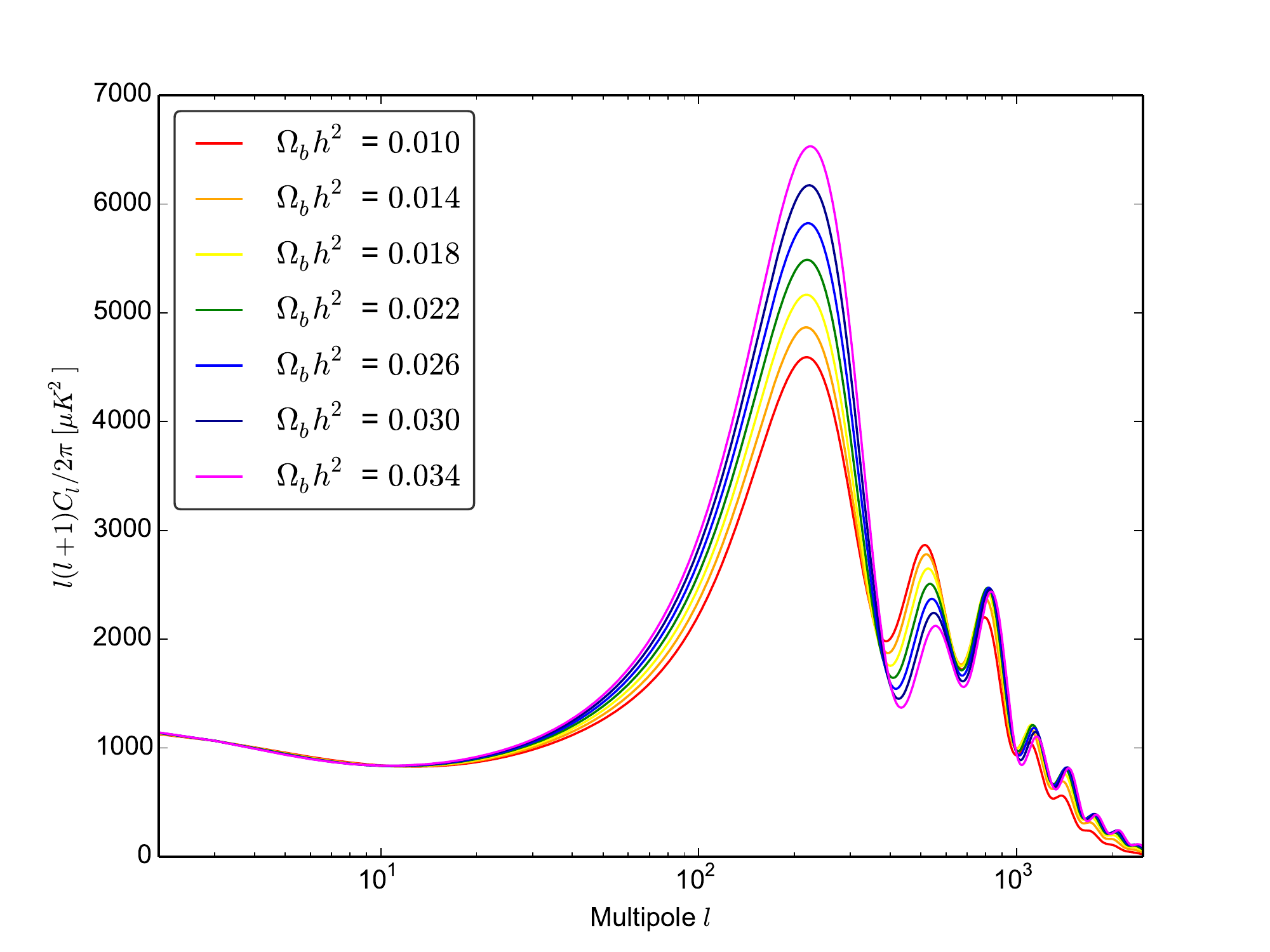}\\
\caption{CMB temperature power spectra, for different values of $\Omega_bh^2$, showing the effect of baryon loading.
We used the CAMB code \cite{Lewis:1999bs} for the fiducial $\Lambda$CDM model given with the following parameters fixed $\Omega_bh^2 = 0.022$, $\Omega_ch^2 = 0.12$, $\Omega_\Lambda = 0.7$, $\tau = 0.1$ and $\Omega_k = 0$.}
\label{Fig2}
\end{figure}

In figure \ref{Fig2} we present how the baryon loading modifies the shape of the CMB temperature power spectrum. 
We vary the baryon density $\Omega_bh^2$ but keep all the other parameters fixed. Because of Friedmann constraint (\ref{Friedmann})
\begin{equation}
 1 = \Omega_b + \Omega_c + \Omega_\Lambda + \Omega_k\;,
\end{equation}
we must therefore let $h$ vary while keeping $\Omega_b$ fixed. Increasing the baryonic density increases the height of the first peak and lowers the second peak. This is due to the fact that, the first peak is a maximum of compression, it is enhanced by a heavier baryon load, while the second peak is a rarefaction peak.

\begin{figure}[ht]
\includegraphics[width=\columnwidth]{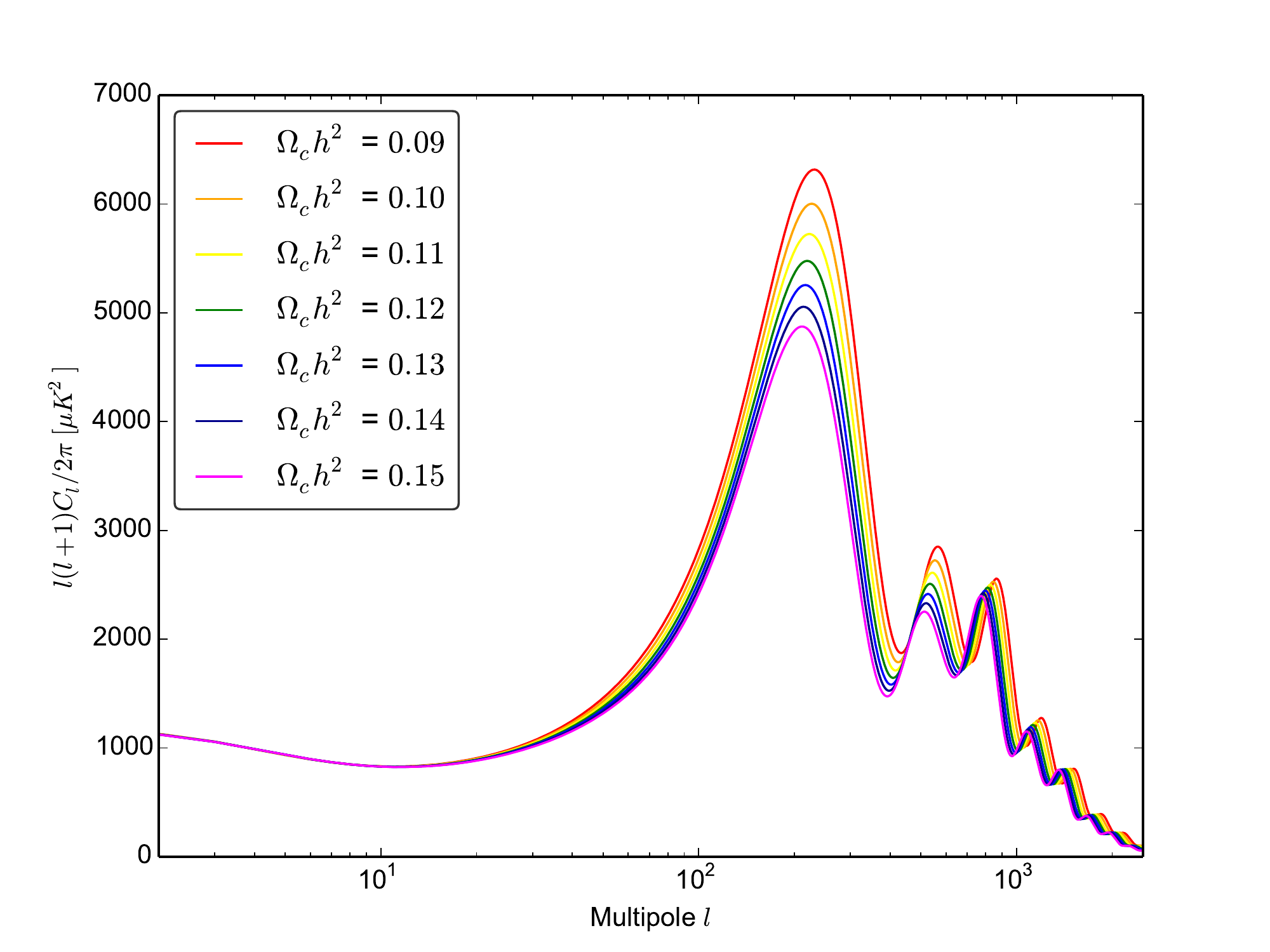}\\
\caption{CMB temperature power spectra, for different values of $\Omega_{c}h^2$, showing the effect of radiation driving.}
\label{Fig2bis}
\end{figure}

In fig. \ref{Fig2bis} we present how the dark matter content modifies the shape of the CMB power spectrum. We use the same strategy as in fig.~\ref{Fig2}, but now we let $\Omega_ch^2$ vary. As one can see, increasing this quantity causes the whole peak structure to decrease in amplitude. Since $\Omega_bh^2$ is fixed, the matter-radiation equivalence epoch takes place at earlier times when increasing $\Omega_ch^2$. This implies weaker radiation driving and a smaller amplitude of oscillations.

There is also another very important feature characterising CMB anisotropies on small scales. On scales comparable to the photon diffusion scale the tight-coupling approximation breaks down. The diffusion comoving scale has the following expression in terms of cosmological parameters~\cite{Hu:2008hd}
 \begin{equation}
 \label{diffusion}
  \lambda_{\rm D} \approx 64.5\mbox{ Mpc }\left(\frac{\Omega_mh^2}{0.14}\right)^{-0.278}\left(\frac{\Omega_bh^2}{0.024}\right)^{-0.18}\;.
 \end{equation}
Photon diffusion translates into a damping of the oscillations, see figure~\ref{Fig3}, and a decay in the correlation, i.e., the angular power spectrum, $C_l$. This is easily understood. On very small scales, below $\lambda_{\rm D}$, cold and hot photons mix thereby averaging to zero the correlation. 

\begin{figure}[ht]
\includegraphics[width=\columnwidth]{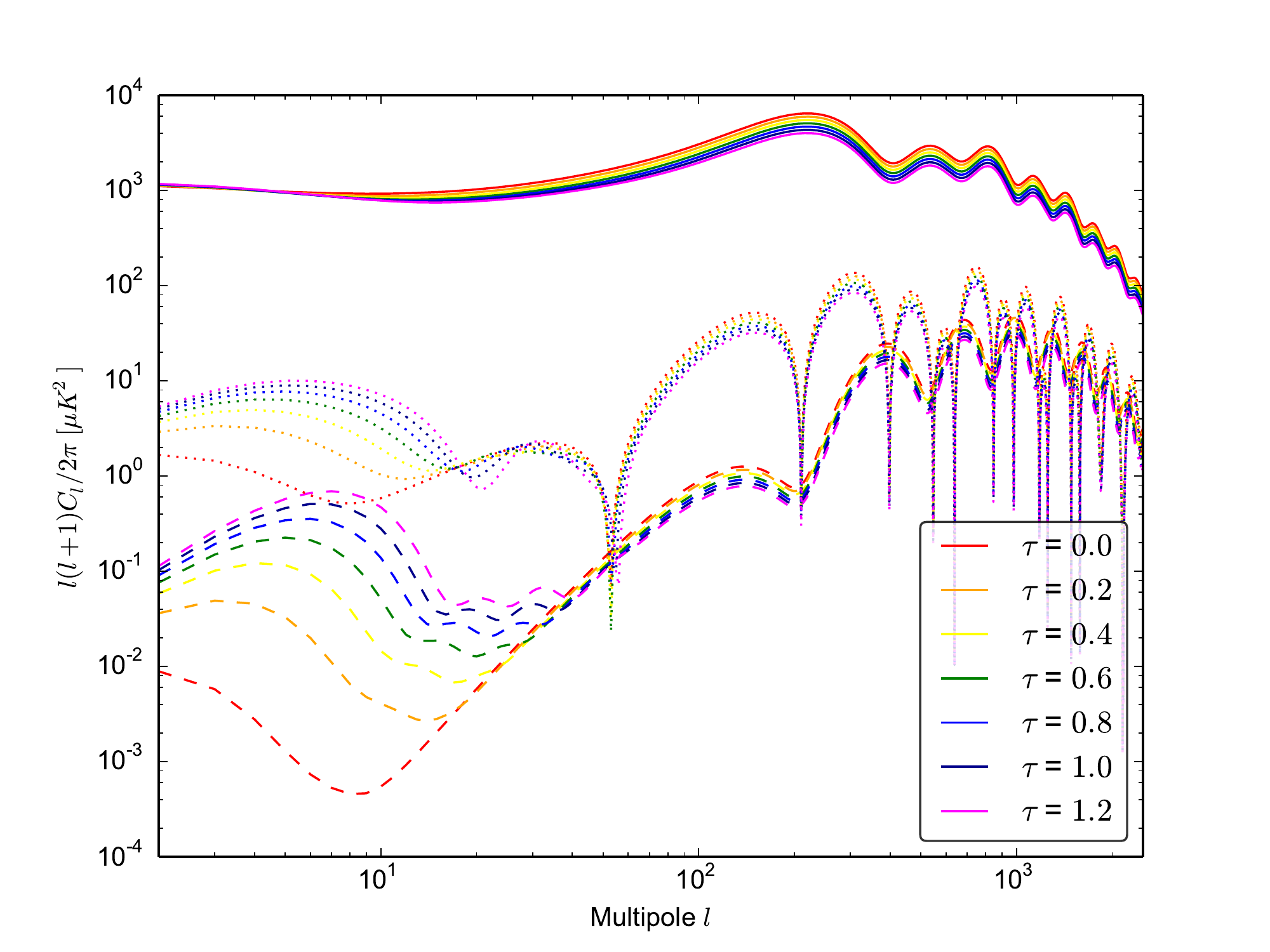}\\
\caption{Effects of reionisation on the temperature (solid lines), EE (dotted lines) and BB (dashed lines) power spectra.}
\label{Fig4}
\end{figure}

Thus far we have assume that all photons last scattered at the time of recombination, but this is an approximation. The absence of a Gunn-Petersen trough (no absorption by neutral hydrogen in quasar spectra) suggests that neutral gas has been reionised by a redshift $z>6$. Reionisation leads to an optically thin ``smog'' between us and recombination with an optical depth $\tau\approx 0.1$, i.e., 10 percent of the photons are scattered again. This reionisation suppresses small-scale anisotropies by rescattering the photons which also tends to average out the temperature anisotropies. The effect of varying the reionisation optical depth is shown in figure~\ref{Fig4}.

\subsection{Parameter constraints from peak structure}

The dependence of the acoustic peak structure on a variety of different physical parameters enables cosmologists to determine these cosmological parameters with unprecedented accuracy from detailed measurements of the CMB sky. This is often referred to as the era of ``precision cosmology''.

From Planck, we have a precise constraint on the angular scale of the first acoustic peak \cite{Ade:2013zuv}:
\begin{equation}
 \theta_* 
 = (1.04148\pm 0.00066)\times10^{-2}\;.
\end{equation}

In flat space this angle is given by $\theta_*=s_*/D_*$, i.e., the ratio between the sound horizon at recombination
\begin{equation}
s_* = \int_0^{\eta_*}c_sd\eta \propto \Omega_mh^2\;,
\end{equation}
and the present horizon distance to recombination
\begin{equation}
 D_* = \eta_0 - \eta_* = \int_0^{z_*}\frac{dz}{H(z)}\propto h^{-1}\;.
\end{equation}
Thus in flat space we find
\begin{equation}
 \label{Omegamh3}
 \theta_* = \frac{s_*}{D_*} \propto \Omega_mh^3 = 0.0959\pm 0.0006\;.
\end{equation}

\begin{figure}[ht]
\includegraphics[width=\columnwidth]{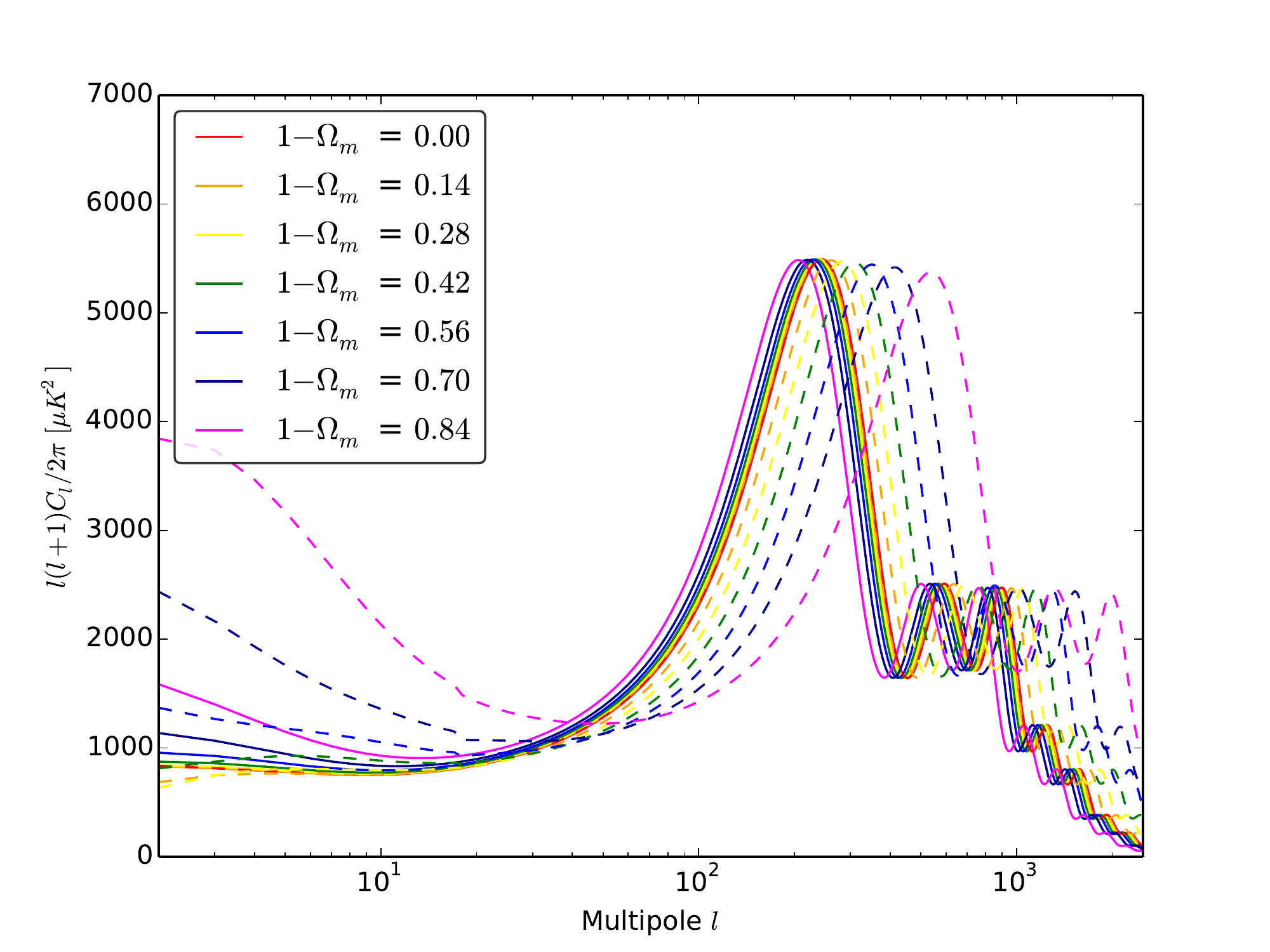}\\
\caption{Effects of varying the cosmological constant (solid lines showing $\Omega_m+\Omega_\Lambda=1$) and spatial curvature (dashed lines showing $\Omega_m+\Omega_k=1$).}
\label{Fig5}
\end{figure}

More generally, in a curved space with curvature radius $R$, we have $\theta_*=s_*/D_A$, see figure \ref{Fig5}.
The angular diameter distance, $D_A$, can be written for $D_*\ll R$ as
\begin{equation}
 D_A = R\sin\left(\frac{D_*}{R}\right) \approx D_*\left(1 + \frac{\Omega_kH_0^2D_*^2}{6}\right)\;,
\end{equation}
%
from which we obtain the bound \cite{Ade:2013zuv}
\begin{equation}
 \Omega_k = -0.0042^{+0.043}_{-0.048}\;.
\end{equation}
As yet there is no evidence for spatial curvature.

We can derive a lot more information from the peak structure. For example, as discussed earlier, the second peak height relative to the first peak is related to the baryon loading. As shown in figure~\ref{Fig2}, this suppresses the even (rarefaction peaks) peaks with respect to the odd numbered peaks (compression peaks). 
From the relative height of the second peak, the following constraint on the the baryon content is found \cite{Ade:2013zuv}:
\begin{equation}
 \Omega_bh^2 = 0.02207\pm0.00033\;.
\end{equation}

The decay of the gravitational potential in the radiation era enhances the third and higher peaks. The less matter there is, the higher the peaks are enhanced, as shown in figure \ref{Fig2bis}. This enables us to put a constraint on the matter density today \cite{Ade:2013zuv}:
\begin{equation}
 \label{Omegamh2}
 \Omega_mh^2 = 0.1423\pm0.0029\;.
\end{equation}
Note that combining the two bounds (\ref{Omegamh3}) and (\ref{Omegamh2}) yields a bound on the Hubble constant (\ref{H0}) directly from the CMB  \cite{Ade:2013zuv}
\begin{equation}
H_0 = 67.3 \pm 1.2\ {\rm km}\ {\rm s}^{-1}\ {\rm Mpc}^{-1} \,. 
\end{equation}

The diffusion length (\ref{diffusion}) and hence the damping tail is then fixed once $\Omega_bh^2$ and $\Omega_mh^2$ are specified in the basic $\Lambda$CDM cosmology. 
Reionisation also suppresses anisotropies at all small angles, above $l = 20$. This is approximately degenerate
with the primordial spectral tilt (\ref{tilt}), but this degeneracy can be broken by polarisation (see next section).
Planck data combined with WMAP polarisation data requires a spectral tilt \cite{Ade:2013zuv}
\begin{equation}
 n = 0.9603\pm0.0073\;.
\end{equation}


\section{Polarisation}


Thomson scattering is due to the motion of charged particles responding to an incident electromagnetic wave. The outgoing radiation emitted by an electron responding to a single incident wave is polarised in the direction of motion of the electron. However for the CMB photons emitted in a given direction at last scattering to have a net polarisation requires the electrons on the last scattering surface to see an anisotropic distribution of incoming photons. We have seen that tight-coupling between electrons and photons leads to an approximately isotropic distribution of photons before recombination, therefore CMB radiation from last-scattering is only weakly polarised, due to a small quadrupole moment caused by small, but finite photon diffusion before last scattering.
However about 10\% of CMB photons are re-scattered long after recombination due to reionisation, when the photon field is anisotropic, and hence reionisation provides a source of polarisation on large angular scales. 


In general we define the polarization tensor in terms of the electric field, $\textbf{E}$~\cite{Hu:2008hd}
\begin{equation}
 \textbf{P}_{ij} \propto \langle \textbf{E}\textbf{E}^\dagger\rangle \propto \left(
 \begin{array}{cc}
 \Theta + Q & U\\
 -U & \Theta - Q
 \end{array}
 \right)
\end{equation}
where $Q$ and $U$ are the two possible states of linear polarization and the 
angle brackets here denote the time average.
Circular polarization is neglected here. 

Since the polarisation state is invariant under a 90$^{\circ}$ rotation, it corresponds to a spin-2 field.
Like the temperature anisotropy, we can decompose the polarisation in any direction into harmonic functions, E and B modes, across the whole sky,  
\begin{equation}
Q(\hat{\textbf{n}})\pm iU(\hat{\textbf{n}}) = \sum_{\ell,m} \left( E_{\ell m}\pm iB_{\ell m} \right) {}_{\pm 2}Y_{\ell m}(\hat{\textbf{n}}) \,,
\end{equation}
where ${}_{\pm 2}Y_{\ell m}(\hat{\textbf{n}})$ are spin-2 spherical harmonics.

\begin{figure}[ht]
\includegraphics[width=\columnwidth]{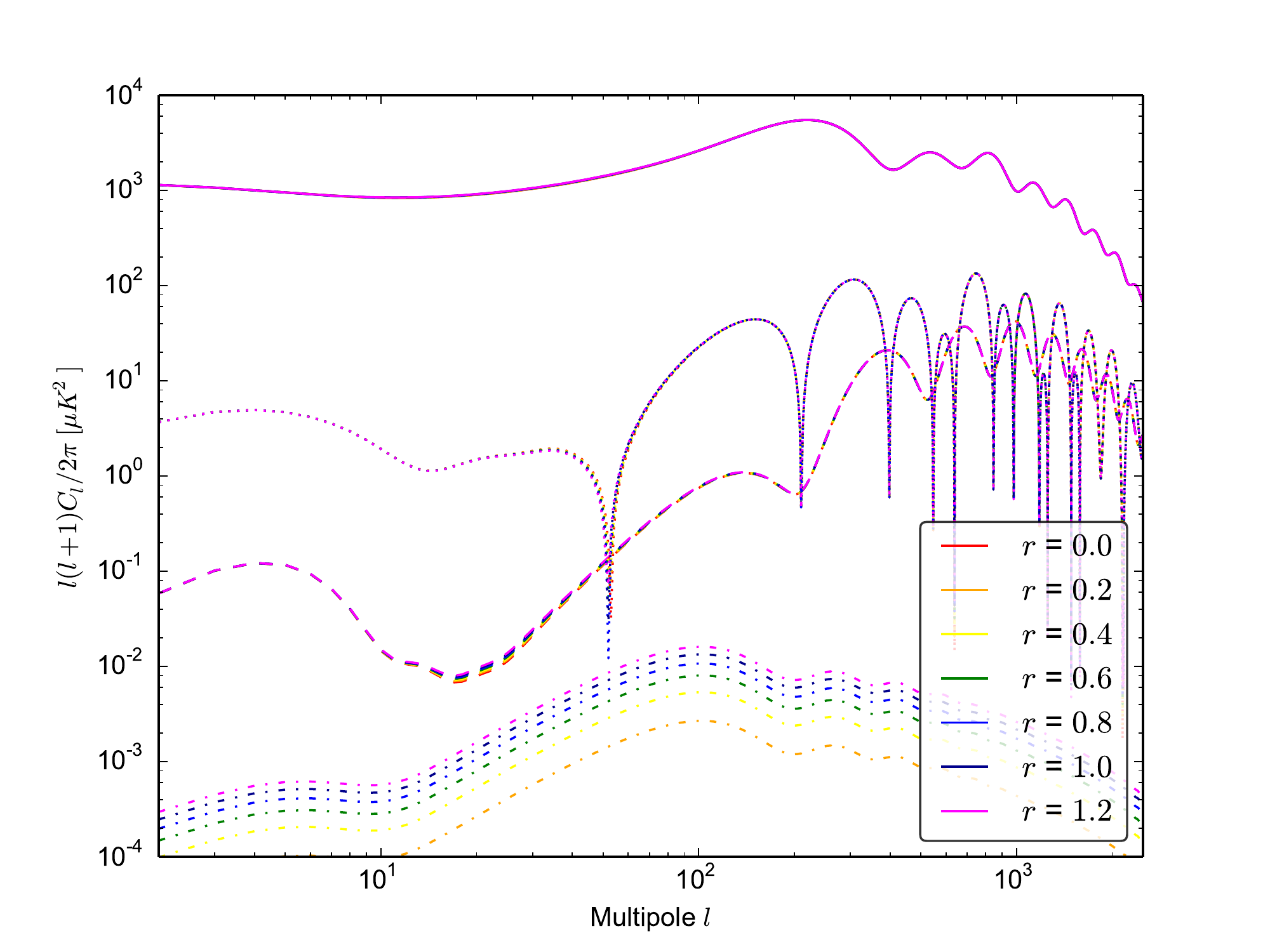}\\
\caption{Effects of varying tensor-to-scalar ratio, $r$. The solid lines (top) correspond to the total intensity (temperature) angular power spectrum. The dashed lines correspond to the E-mode power spectrum and the dotted lines correspond to the temperature-E-mode cross-correlation. These are dominated by scalar modes. The dot-dashed lines show the B-mode angular power spectrum for different values of the primordial tensor-to-scalar ratio, $r$.}
\label{Fig6}
\end{figure}

Scalar perturbations are longitudinal wave-modes where the inhomogeneities which give rise to polarised radiation are in the same direction as the wave propagates. This symmetry ensures that only E-mode polarisation is generated by scalar perturbations at linear order, corresponding to polarisation parallel or perpendicular to density gradients. Figure \ref{Fig6} shows the angular power spectrum for E-modes alongside the temperature power spectrum, and their cross-correlation. It shows the E-mode polarisation ``bump'' at large angular scales ($\ell<20$) which can be used as a sensitive measure of reionisation and optical depth $\tau$.

B-mode polarisation can only be generated at first order by transverse waves, i.e., vector or tensor perturbations. Initial vector perturbations decay at linear order in an expanding cosmology, and are completely absent in inflation driven by scalar fields. However tensor perturbations correspond to free perturbations of the metric, i.e., gravitational waves.

Consider a spatial metric perturbation:
\begin{equation}
 h_{ij}(\textbf{x},\eta) = \int\frac{d^3\textbf{k}}{(2\pi)^{3/2}}e^{i\textbf{k}\cdot\textbf{x}}\left[h_\textbf{k}e_{ij}(\textbf{k}) + \bar h_\textbf{k}\bar e_{ij}(\textbf{k})\right]\;.
\end{equation}  
Tensor perturbations must be transverse and traceless, i.e.,
\begin{equation}
 k^ie_{ij} = 0\;, \qquad k^i\bar{e}_{ij} = 0\;, \qquad g^{ij}e_{ij} = 0\;, \qquad g^{ij}\bar{e}_{ij} = 0\;.
\end{equation}
This leaves only two independent degrees of freedom, which are the two polarisation modes characterising a gravitational wave. 

The Einstein equations give the following evolution equation for the gravitational wave amplitude
\begin{equation}
 \ddot h_k + 3H\dot h_k + \frac{k^2}{a^2}h_k = 0\;,
\end{equation}
i.e., a wave equation for a massless field, including a damping term due to the expansion of the Universe. Quantum vacuum fluctuations in massless fields during slow-roll inflation generate an almost scale-invariant spectrum of primordial perturbations on super-Hubble scales $k < aH$. Therefore, we can predict a primordial power spectrum for gravitational waves from inflation:
\begin{equation}
 \mathcal{P}_T(k) = 2\frac{4\pi k^3}{(2\pi)^3}\langle h h^\dagger\rangle \approx 2\frac{32\pi}{M_{Pl}^2}\left.\frac{H^2}{2\pi}\right|_{k = aH}\;,
\end{equation}
where the average is the vacuum expectation value, classically promoted to a variance. 
The power spectrum is thus directly determined by the Hubble parameter, $H$, during inflation.


It is customary to introduce the tensor-to-scalar ratio:
\begin{equation}
 r \equiv \frac{\mathcal{P}_T}{\mathcal{P}_\zeta}
 \;.
\end{equation}
From Planck constraints on the temperature power spectrum we obtain \cite{Ade:2013zuv}, $r < 0.11$, and the energy scale at the end of inflation is constrained to be $V < 2\times 10^{16}$ GeV.
Nonetheless there is considerable interest in searching for B-mode polarisation in the CMB, either from future satellite experiments or dedicated ground-based experiments, as a direct signal of primordial gravitational waves. If an almost scale invariant, Gaussian distribution of primordial gravitational waves were discovered then these would surely be strong evidence for inflation, and our first evidence for the quantum nature of gravity.


\section{The next frontier in CMB theory}

The observed CMB sky is remarkably uniform with temperature variations less than one part in $10^4$. These lectures have reviewed the analysis of the homogeneous ``background'' CMB sky and anisotropies in the CMB temperature and polarisation, modelled using first-order perturbation theory. As the precision of CMB experiments improves the next challenge in CMB theory may be to accurately model non-linearity in the CMB anisotropies, both at last scattering and along the subsequent line-of-sight.

Non-linear interactions lead to departures from Gaussianity, evident in the CMB bispectrum, i.e., a non-zero correlation between different spherical harmonics. 
Primordial non-Gaussianity is often described in terms of the dimensionless nonlinearity parameter \cite{Bartolo:2004if}
\begin{equation}
 f_{NL} \equiv \frac{B_\zeta(k_1,k_2,k_3)}{P_\zeta(k_1)P_\zeta(k_2) + P_\zeta(k_2)P_\zeta(k_3) + P_\zeta(k_1)P_\zeta(k_3)} \; ,
\end{equation}
i.e., the primordial bispectrum, $B_\zeta(k_1,k_2,k_3)$, relative to the square of the power spectrum. This $f_{NL}\sim1$ corresponds to a primordial bispectrum $B_\zeta(k_1,k_2,k_3)\sim10^{-18}$.
It is related to the three-point correlation function, which is identically vanishing in the Gaussian case, along with all the odd-order correlation functions. In general $f_{NL}$ defined in this way is a scale- and shape-dependent function of the three wave numbers, but for local-type non-Gaussianity $f_{NL}$ is a constant parameter \cite{Wands:2010af}.

Non-linear interactions also lead to important conceptual differences from the simple assumptions valid in linear theory. Small but non-zero vector and tensor perturbations, hence B-mode polarisation, are generated at second-order from first-order scalar perturbations \cite{Mollerach:2003nq,Fidler:2014oda}. Also, second order anisotropies in the the photon distribution can no longer be described simply by a black-body spectrum with anisotropic temperature. Second-order effects lead to anisotropic spectral distortions. For example, the angular power spectrum of the Compton y-distortion could provide a powerful tracer of reionisation \cite{Pitrou:2009bc}.

Weak lensing along the line of sight is already an important nonlinear effect seen in current data, and needs to be taken into account in parameter estimates using Planck data.
It is caused by many small-angle deflections by non-linear structures along the line of sight. This redistributes power in the small-scale angular power spectrum, smoothing out peaks at high $\ell$ in the power spectrum. This anisotropic lensing of the small scale power in the CMB has been used by the Planck team to create a map of the lensing potential and hence a measure of the integrated mass distribution along the line of sight \cite{Ade:2013tyw}.

CMB weak-lensing also provides a non-zero contribution the angular bispectrum.
Gravitational lensing leads to a second order anisotropy which is correlated with the integrated Sachs-Wolfe effect along the line of sight \cite{Goldberg:1999xm, Seljak:1998nu, Lewis:2011fk}. 
This contributes a significant bias to estimates of the non-linearity parameter, and $f_{\rm NL} \approx 7$ has been seen in the Planck analysis.
After subtracting this effect, Planck measurements remain consistent with vanishing primordial non-Gaussianity, $f_{NL} = 2.7\pm5.8$ \cite{Ade:2013ydc}.

\begin{figure}[ht]
\includegraphics[width=\columnwidth]{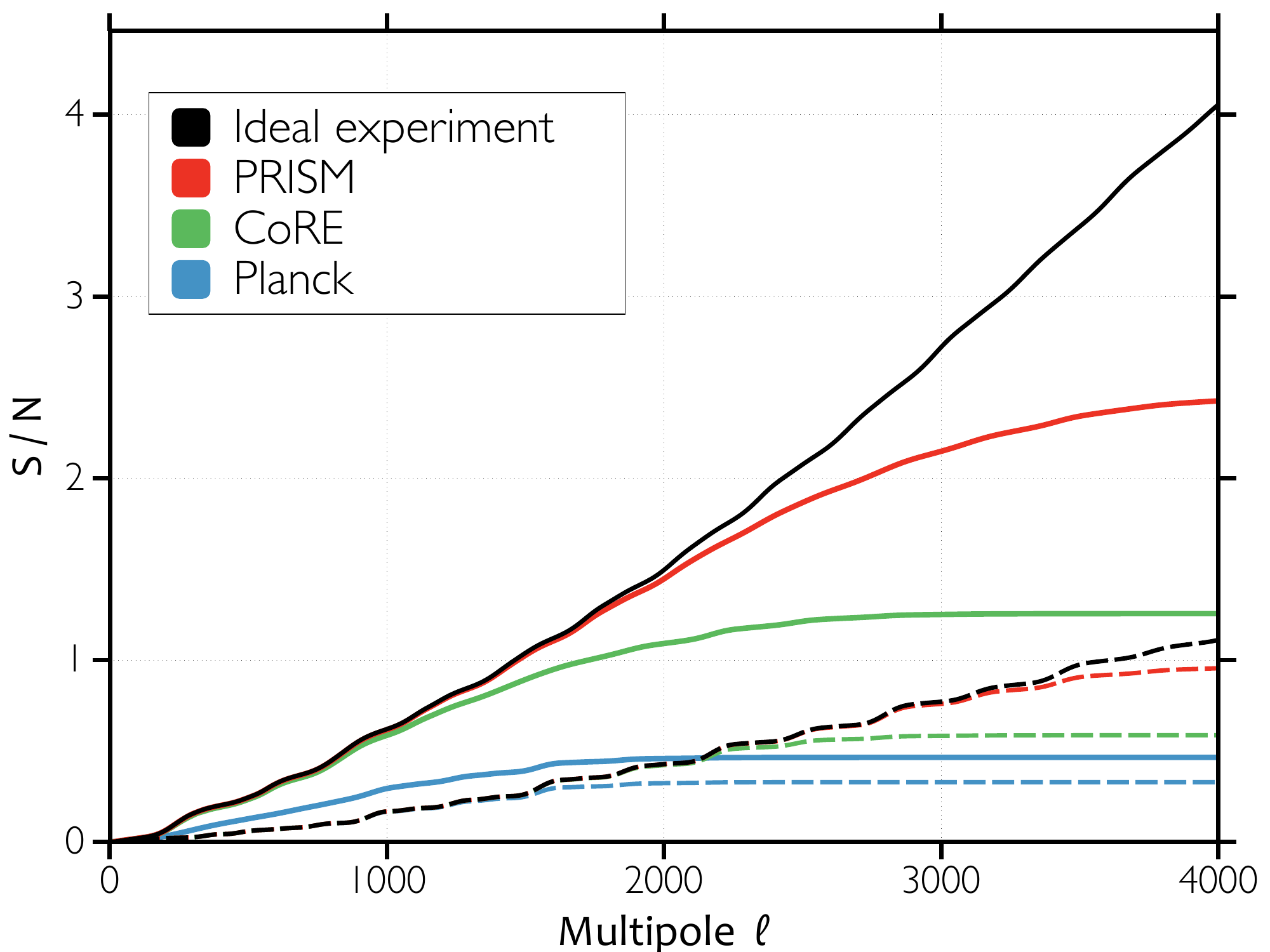}
\caption{Signal-to-noise ratio of the intrinsic bispectrum signal as a function of maximum multipole for Planck and proposed future satellite missions using temperature only (dashed lines) or temperature and polarisation (solid lines). Figure reproduced from \cite{Pettinari:2014iha}. Copyright (2014) by The American Physical Society.}
\label{Fig7}
\end{figure}

Most current bounds on primordial non-Gaussianity are based upon theoretical templates based on non-linear modelling of inflationary (or alternative) models, as non-Gaussian initial conditions for standard, linear Boltzmann codes such as CAMB \cite{Lewis:1999bs} or CLASS \cite{Lesgourgues:2011re}. However the process of decoupling and the Sachs-Wolfe effect on temperature (and spectrum) anisotropies is in reality a non-linear process. As bounds on primordial non-Gaussianity become tighter we also need templates for the intrinsic non-Gaussianity expected from non-linear physics at recombination. There are now second-order general relativistic Boltzmann codes which have been developed  \cite{Huang:2012ub, Pettinari:2013he, Su:2014tga} building on pioneering early work \cite{Pitrou:2010ai}. 
Intrinsic non-Gaussianity at last-scattering provides a small bias, $f_{NL}\simeq1$, which remains below the observational uncertainty of current experiments. However we are now in the position to be able to build a template for the intrinsic non-linearity at recombination which could be a target for future all-sky (hence space-based) missions, as shown in figure~\ref{Fig7}. This would make novel tests of physical process at last-scattering, e.g., gravitational wave production from density waves at second order.

\section{Outlook}

The standard $\Lambda$CDM cosmology provides a remarkably successful base model, able to explain many detailed features of the CMB revealed over the past fifty years by a series of ground-, balloon- and space-based experiments. In particular the Sachs-Wolfe plateau at large angular scales, the series of acoustic peaks in the angular power spectrum, and the damping tail at small angular scales can be described by this model with just six cosmological parameters: the Hubble constant, the baryon and matter densities, the reionisation optical depth, and the amplitude and tilt of primordial perturbations.
These six parameters are increasingly tightly constrained in the new era of precision cosmology, and bounds are set to become ever tighter, especially through new combinations with other data sets, such as high redshift galaxy surveys and HI (neutral hydrogen) survey data.
The framework already successfully accommodates new observational discoveries such as the effect of weak lensing now seen in the CMB power spectrum and bispectrum.

Nonetheless there is no reason to believe this is the final theory of cosmology. 
Even the simplest, single-field models of inflation in the very early universe make predictions for additional features in the primordial perturbations, including a spectrum of tensor (gravitational wave) perturbations and small but finite running of the scalar spectral index. 
Many inflation models make further predictions including primordial isocurvature perturbations and/or non-Gaussianity. 
Any of these additional parameters would radically change our views about the likely mechanisms generating primordial structure.
There are many additional cosmological parameters possible, including additional particle species and/or interactions, but there is no clear evidence yet requiring any more than the six basic parameters.

The present theoretical framework now being constrained by data was established in the 1970s and 80s well before the golden age of CMB experiments was begun by the COBE satellite results. Work now in progress will set new theoretical challenges for future experiments. Ground-based experiments are currently targetting CMB polarisation from weak-lensing and the elusive B-mode signature of primordial gravitational waves. Future space-based experiments are likely to focus on polarisation, spectral distortions and/or non-Gaussianity.
The CMB will remain a key testing ground for cosmological theory for many years to come.

\section*{Acknowledgements}

DW is grateful to the organisers of the second Jos\'e Pl\'inio Baptista school for their warm hospitality. The authors are grateful to Rob Crittenden for helpful comments. This work is supported by STFC grants ST/K00090/1 and ST/L005573/1.

\bibliographystyle{unsrt}
\bibliography{BiblioCMB}

\end{document}